\documentclass[sigconf]{acmart}

\pdfoutput=1
\usepackage{fancyhdr}
\usepackage[keeplastbox]{flushend}

\usepackage{amsmath,amsfonts}
\usepackage{algorithmic}
\usepackage{graphicx}
\usepackage{textcomp}
\usepackage{xcolor}
\usepackage[font={small, bf}, labelfont={small, bf}]{caption}

\usepackage{xspace}
\usepackage{wrapfig}
\usepackage{comment}
\usepackage{tabularx}
\usepackage{listings}
\usepackage{diagbox}
\usepackage{subcaption}
\usepackage{xspace}
\usepackage{svg}
\usepackage{array}
\usepackage{xspace}

\hypersetup{
    bookmarks=true,
    breaklinks=true,
    letterpaper=true,
    colorlinks,
    citecolor=blue,
    linkcolor=blue,
    urlcolor=blue
}

\AtBeginDocument{
  }

\copyrightyear{2023}
\acmYear{2023}
\setcopyright{acmlicensed}\acmConference[MICRO '23]{56th Annual IEEE/ACM International Symposium on Microarchitecture}{October 28-November 1, 2023}{Toronto, ON, Canada}
\acmBooktitle{56th Annual IEEE/ACM International Symposium on Microarchitecture (MICRO '23), October 28-November 1, 2023, Toronto, ON, Canada}
\acmPrice{15.00}
\acmDOI{10.1145/3613424.3623791}
\acmISBN{979-8-4007-0329-4/23/10}

\newcommand{\specname}{TeAAL\xspace}
\newcommand{\fullname}{Tensor Algebra Accelerator Language}

\newcolumntype{P}[1]{>{\centering\arraybackslash}p{#1}}
\newcolumntype{M}[1]{>{\centering\arraybackslash}m{#1}}

\begin{document}

\title[\specname: A Declarative Framework for Modeling Sparse Tensor Accelerators]{\specname: A Declarative Framework \\ for Modeling Sparse Tensor Accelerators}

\author{Nandeeka Nayak}
\email{ndnayak2@illinois.edu}
\affiliation{
  \institution{University of Illinois Urbana-Champaign}
  \city{Urbana}
  \state{Illinois}
  \country{USA}
}

\author{Toluwanimi O. Odemuyiwa}
\email{todemuyiwa@ucdavis.edu}
\affiliation{
  \institution{University of California, Davis}
  \city{Davis}
  \state{California}
  \country{USA}}

\author{Shubham Ugare}
\email{sugare2@illinois.edu}
\affiliation{
  \institution{University of Illinois Urbana-Champaign}
  \city{Urbana}
  \state{Illinois}
  \country{USA}
}

\author{Christopher W. Fletcher}
\email{cwfletch@illinois.edu}
\affiliation{
  \institution{University of Illinois Urbana-Champaign}
  \city{Urbana}
  \state{Illinois}
  \country{USA}
}

\author{Michael Pellauer}
\email{mpellauer@nvidia.com}
\affiliation{
  \institution{NVIDIA}
  \city{Westford}
  \state{Massachusetts}
  \country{USA}
}

\author{Joel S. Emer}
\email{emer@csail.mit.edu}
\affiliation{
  \institution{MIT/NVIDIA}
  \city{Cambridge}
  \state{Massachusetts}
  \country{USA}
}

\renewcommand{\shortauthors}{Nayak et al.}

\begin{abstract}

Over the past few years, the explosion in sparse tensor algebra workloads has led to a corresponding rise in domain-specific accelerators to service them.
Due to the irregularity present in sparse tensors, these accelerators employ a wide variety of novel solutions to achieve good performance.
At the same time, prior work on design-flexible sparse accelerator modeling does not express this full range of design features, making it difficult to understand the impact of each design choice and compare or extend the state-of-the-art.

To address this, we propose \specname: a language and simulator generator for the concise and precise specification and evaluation of sparse tensor algebra accelerators.
We use \specname to represent and evaluate four disparate state-of-the-art accelerators---ExTensor, Gamma, OuterSPACE, and SIGMA---and verify that it reproduces their performance with high accuracy.
Finally, we demonstrate the potential of \specname as a tool for designing new accelerators by 
showing how it can be used to speed up vertex-centric programming accelerators---achieving $1.9\times$ on BFS and $1.2\times$ on SSSP over GraphDynS.

\end{abstract}

\maketitle

\section{Introduction}

Sparse tensor algebra workloads have exploded in popularity over the past few years, with applications ranging from deep learning~\cite{sze:2020:epo, mahmoud:2020:tde, albericio:2016:cin} to graph algorithms~\cite{graph_clustering,nagasaka2019performance,graphblas,triangle_counting_tensor_algebra,tallskinny0,tallskinny1} to physical simulations~\cite{vande:2012:lss,wilhelm:2016:lsc,hutter:2014:cas}.
This surge has been accompanied by a corresponding rise in proposals for custom hardware to service common sparse  
kernels, e.g., sparse matrix multiply~\cite{gamma,extensor,outerspace,sigma,sparch,eie,SCNN}.

While these accelerators have the potential to provide dramatic speedup over the best CPU and GPU algorithms, they take significant effort and space to describe, refine,
and evaluate.
Specifically, sparse accelerators are typically described either with RTL or a block diagram and an accompanying natural language description.
The former is verbose and often difficult to comprehend, while the latter is imprecise and often incomplete.
Neither makes it easy to model and evaluate the impact of proposed design changes.

The goal of our work is to ameliorate these issues.
That is, to enable the precise and concise specification of sparse tensor algebra accelerators, thereby providing a basis for describing, modeling, evaluating, comparing, and extending proposed designs.

We draw inspiration from existing practice in \emph{dense} tensor algebra accelerator design.
Here, a number of tools support concise, precise specifications and the derivation of efficient models~\cite{timeloop, interstellar, maestro}.
To simplify specification, these tools follow the model proposed by Halide~\cite{halide} and support separately providing a target \emph{algorithm} (i.e., a functional description of the problem, such as an equation in Einstein summation (Einsum) notation~\cite{einsum}) and a \emph{mapping}, expressing when and where in the processor each action (e.g., compute or storage access) occurs~\cite{eyeriss}. 
These, together, correspond to a \emph{mapped representation} (e.g., a loop nest), describing both the algorithm and how it is executed.
A \emph{model of the target platform} then evaluates the mapped representation to produce metrics like performance and energy.

Yet, 
dense tensor accelerator modeling techniques  
cannot support the sparse case.
This is due to the 
novel complexity that arises when trying to efficiently orchestrate and compute on irregularly sparse data.
For example, one can accurately model dense kernels using a few summary statistics (like tensor shapes).
However, such summary statistics cannot capture the variety in sparsity distributions present in real-world tensors.
Indeed, sparse accelerators regularly employ specialized algorithms and mechanisms to cope with irregular (often low) data reuse, myriad compression formats, additional meta-computation (e.g., intersection), and more.
For example, the OuterSPACE accelerator~\cite{outerspace} splits sparse-sparse matrix multiply (SpMSpM) into several phases that respectively produce, sort, and consume an array of linked lists representing partial products.
Gamma~\cite{gamma} executes the same kernel with two stages that are connected with a high-radix hardware merger to process the data efficiently.
SIGMA~\cite{sigma} uses yet a third strategy (irregularly filling a PE array with only non-zero data).  And so on.
All of the above stems from irregular sparsity, which obviously does not manifest when data is dense.

Likewise, existing tools for modeling \emph{sparse} tensor algebra accelerators do not fully overcome challenges arising from irregular sparsity.
For example, STONNE~\cite{stonne, flexagon} supports only the SpMSpM kernel,
and even then, only six pre-selected mappings for that kernel.
Sparse\-loop~\cite{sparseloop_micro} can model an accelerator describable in a single deep loop nest.
As we will show, this is insufficient to express SIGMA, OuterSPACE, and Gamma. 
Additionally, Sparseloop uses abstract distribution functions to model sparsity, rather than precisely modeling the behavior of actual input sets.
While better than using \emph{just} shape-based information (like dense modeling), we show how this approach still results in degraded modeling accuracy (Section~\ref{sec:eval:validation}). 

\smallskip
\noindent \textbf{Our contribution.}
We provide a basis for specifying sparse tensor algebra accelerators by showing how recent designs
can be expressed as \emph{cascades (directed acyclic graphs or DAGs) of mapped Einsums}
and \emph{content-preserving transformations} on the 
constituent
tensors in those Einsums.
For example, both OuterSPACE and Gamma can be described by rewriting the Einsum for matrix multiply into several, dependent Einsums.
In this abstraction, the sort/merge operations in those designs are described as \emph{reordering the dimensions} of an intermediate tensor to improve locality, 
while the differences between the two operations are captured in how each Einsum is mapped.
We show how the design choices of other accelerators can likewise be described in terms of a small set of categories of operations, e.g., splitting/combining dimensions of a tensor while preserving its contents.

Based on the above abstraction, we propose \specname (for \emph{\fullname}), a novel declarative language, simulator generator, and performance model 
that enables precise design specification and modeling of sparse tensor algebra accelerators.
The \specname simulator generator takes accelerators described as mapped Einsum cascades and produces an imperative-style \emph{intermediate representation (IR)} that describes tensor transformations as primitive operations on tensors represented as \emph{fibertrees}~\cite{sze:2020:epo}.
It then uses 
implementation-level specifications (e.g., describing the tensor formats)
to augment the IR
to produce an accurate, validated performance model that processes real tensors.

Although attributes such as language expressivity and conciseness are difficult to quantify, as part of a study to validate \specname's fidelity, we 
write the \specname Einsum and mapping specifications of four recent (and disparate) sparse tensor algebra accelerator proposals (OuterSPACE~\cite{outerspace}, ExTensor~\cite{extensor}, Gamma~\cite{gamma}, and SIGMA~\cite{sigma}) in less than a page (see Figures~\ref{fig:outerspace:spec} and \ref{spec:accelerators}), with each specification taking 
$\sim30$ lines.
We verify that the models generated for each of these accelerators reproduce the designs' original published performance results (given the same input data sets) with high accuracy.

We view our primary contribution to be the accurate specification and modeling of sparse tensor algebra accelerators.
That said, we also show how \specname can be applied in adjacent domains and be used to explore optimization opportunities for accelerators in those domains.
Specifically, we use \specname to describe Graphicianado~\cite{graphicionado} and GraphDynS~\cite{graphdyns}, which accelerate vertex-centric programming (a popular paradigm for graph algorithms), and demonstrate how to improve these designs by making point changes to their \specname specifications.

Taken as a whole, we feel \specname constitutes a significant advance over state-of-the-art practices in modeling and evaluating sparse tensor accelerators.
Using \specname, one is able to describe a design using a precise, unified set of abstractions, supporting qualitative and quantitative comparison to other designs and enabling the exploration of the impact of a series of design changes.
By contrast, standard practice today is to rely on English descriptions/figures and bespoke simulators, which suffer along all of these dimensions.
Although prior works exist for exploring the space of sparse accelerator designs (e.g., Sparseloop~\cite{sparseloop_micro}), they have more limited expressivity.
For example, Sparseloop is only able to represent one of the six accelerators that we show results for.

To summarize, we make the following contributions:
\begin{itemize}
\item We show how modern sparse tensor algebra accelerator features can be represented using cascades of mapped Einsums and content-preserving transformations on those Einsums' constituent tensors.
Based on this abstraction, we propose the \specname specification language for concisely and accurately specifying the design of sparse tensor algebra accelerators.
\item We propose and design a simulator generator
that transforms \specname specifications into an imperative-style IR that performs operations on fibertrees and lowers that IR to an accurate performance model of the specified design that processes
real sparse tensor inputs.
\item We validate \specname's accuracy in terms of modeling memory traffic, performance, and energy with respect to the reported results of
four state-of-the-art accelerators.
\item We demonstrate the potential of \specname as a tool for accelerator design by using it to speed up accelerators for vertex-centric programming---by $1.9\times$ on BFS and $1.2\times$ on SSSP over GraphDynS~\cite{graphdyns}.
\end{itemize}

Beyond the artifact (Appendix~\ref{sec:app_ae}), we have made the source code for \specname available at \url{https://github.com/FPSG-UIUC/teaal-compiler}.

\section{Background and Motivation}
\label{sec:background}

We review key attributes of sparse tensor algebra and outline the design decisions typically made by sparse tensor accelerators. 
We highlight the difficulties with informal comparisons between accelerators and motivate the need for a precise, formal specification.

\subsection{Tensors and Fibertrees}
\label{sec:background:tensors}

\begin{figure}[t]
\centering
\includegraphics[width=0.85\columnwidth]{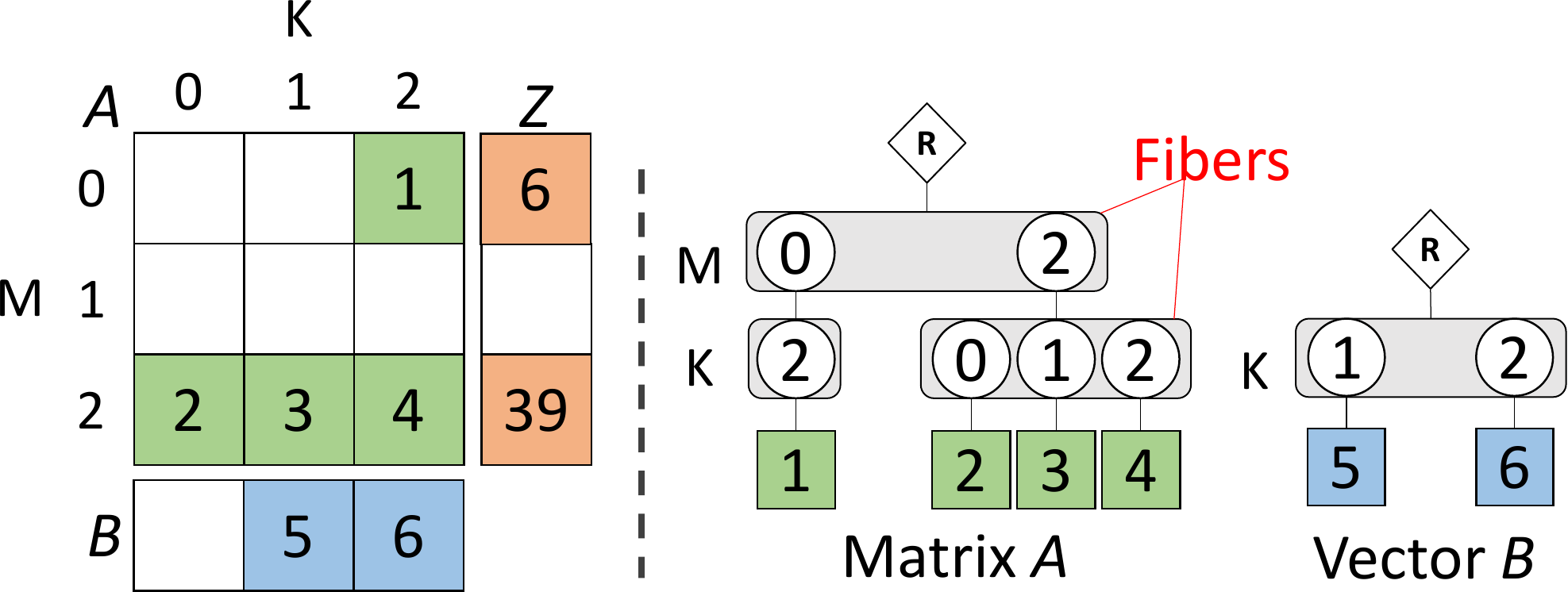} 
    \vspace{-0.05in}
  \caption{\small Sparse matrix-vector multiplication and corresponding fibertree\@ representations\@.}
  \label{fig:background:fibertree}
  \vspace{-0.15in}
\end{figure}

In this paper, an $N$-\emph{tensor} is a multidimensional array with $N$ dimensions. 
For example, a 0-tensor is a scalar, a 1-tensor is a vector, and a 2-tensor is a matrix. Figure \ref{fig:background:fibertree} shows a 2-tensor, $A$, with dimensions $M$ and $K$. 
Using the terminology from Sze et al.~\cite{sze:2020:epo}, we describe a tensor's attributes:
\begin{itemize}
    \item A \emph{rank} refers to an axis/dimension in the tensor. A matrix has two ranks, often described as rows and columns. 
    \item A \emph{point} is a 
    logical location within a tensor that contains a scalar \emph{value}. A point is identified by an $N$-tuple of \emph{coordinates} with one coordinate for each rank in the tensor.
    We denote the tensor $A$'s element at point $(m,k)$ as $A_{m,k}.$
\end{itemize}

Mathematically, tensors have no notion of sparsity or compression format (e.g., CSR).
To avoid getting bogged down in the numerous details of various formats, we leverage the following abstractions proposed in Sze et al.~\cite{sze:2020:epo}:

\begin{itemize}
    \item A \emph{fibertree} represents a tensor as a 
    tree, with each level corresponding to a labeled rank in the tensor. Tensor $A$ in Figure~\ref{fig:background:fibertree} has ranks $M$ and $K$. 
    
    \item The order of levels in a fibertree reflects its \emph{rank order}, denoted $[M, K]$ in our example.
    The rank order list read left-to-write corresponds to the fibertree's ranks read top-to-bottom in the tree.
    
    \item Every level contains one or more \emph{fibers}.  A \emph{fiber} is the set of elements sharing all coordinates in all higher levels of the tree.
    Fibers are more precise than ``rows'' or ``columns,'' because they naturally extend to $N$-tensors.
    
    \item Each \emph{element} in the fiber is a coordinate/payload pair, where the \emph{payload} is a scalar value when it is at a leaf or a reference to a fiber when it is an intermediate node.

    \item The \emph{shape of a fiber} is the the set of legal values the coordinates in that fiber can take on, where an integer shape means the open interval from zero to that integer.
    The \emph{shape of a rank} is the union of the shapes of all fibers in that rank.
    The \emph{shape of a tensor} is the list of shapes of each of the ranks in rank order.
    
\end{itemize}

\begin{figure}
  \includegraphics[width=\columnwidth]{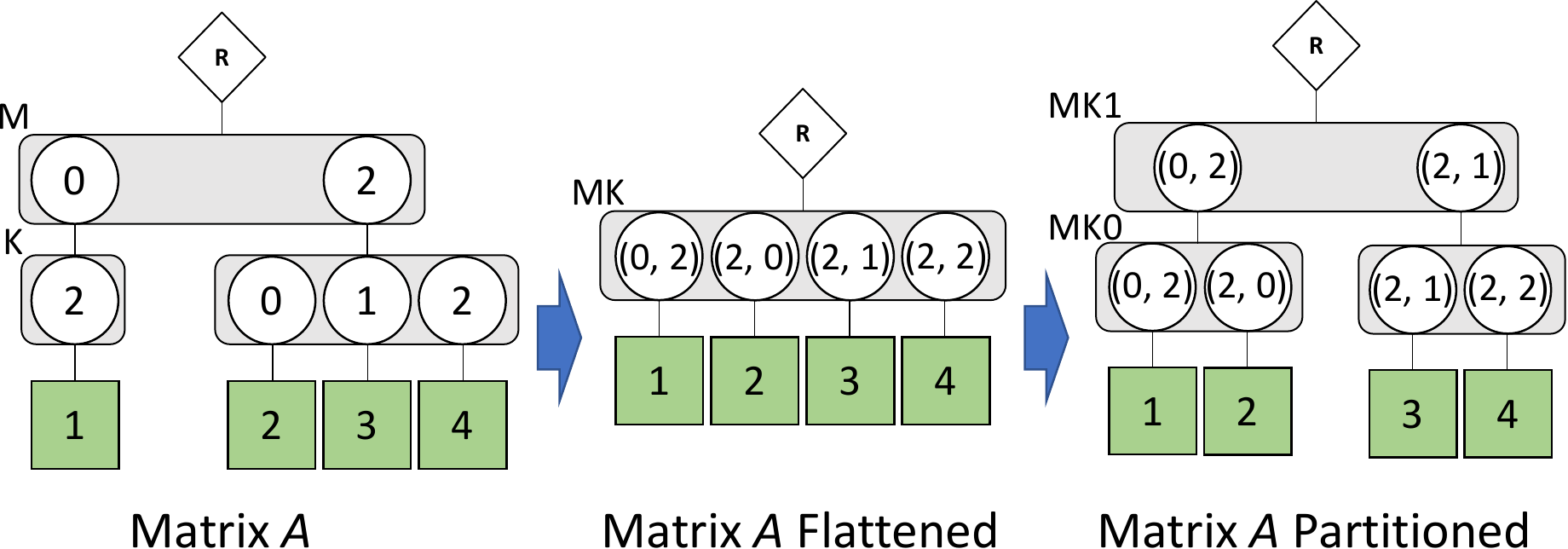}    \vspace{-0.2in}
  \caption{\small Flattening then partitioning ranks $M$, $K$ of tensor $A$ (Fig. \ref{fig:background:fibertree}).
  }
  \label{fig:uniform_occupancy}
  \label{fig:insights:partitioning}
  \vspace{-0.15in}
\end{figure}

One advantage of fibertrees is that they naturally handle both dense and sparse tensors (i.e., tensors where a number of points are zero).
A dense tensor's fibertree has every coordinate present in the entire shape (i.e., a complete tree).
On the other hand, a sparse tensor's fibertree can omit 
all elements with empty payloads (either zero values or empty fibers).
The semantics of operations on fibers and fibertrees remain the same in both cases.
Note that fibertrees are just an abstraction we use to describe operations on tensors.
To model a specific design, all fibertrees are lowered to concrete representations, like CSR or COO (see Section~\ref{sec:model:specs:format}).
In this work, we use fibertrees both to categorize the space of design choices (Section~\ref{sec:insights}) and as an IR during performance modeling (Section~\ref{sec:model}).

The fibertree abstraction also supports a number of 
transformations that change the fibertree corresponding to a tensor:
\begin{itemize}
    \item A \emph{rank flattening}, demonstrated in the first transformation in Figure~\ref{fig:insights:partitioning}, combines two ranks together into a single rank.
    After flattening, the coordinates are tuples of the coordinates in the original fibers that reference a payload from the original lower rank.
    \item A \emph{rank partitioning}, demonstrated in the second transformation in Figure~\ref{fig:insights:partitioning}, separates a rank into two ranks.
    The coordinates of the new upper rank denote the first legal coordinate in the fiber below.
    \item A \emph{rank swizzle} changes the fibertree's rank order (i.e., reorders the levels of the fibertree).
\end{itemize}

An important insight in our work is that
many sparse accelerator behaviors can be viewed as one or more of these transformations (Section~\ref{sec:insights}).

\begin{table*}[h!]
\hfill
\centering
\footnotesize
\caption{\label{motivation:accelerators} \small Comparison of selected sparse tensor accelerator hardware proposals. TeAAL specifications increase both the precision and formalism of such comparisons, and enable automatic generation of performance/energy models.}
\vspace{-0.05in}
\begin{tabular}{ p{1.9cm} | p{.5cm} | p{5.7cm} | p{8.2cm} }
\textbf{Accelerator} & \textbf{Year} & \textbf{Mapping Approach} & \textbf{Architectural Focus} \\ \hline \hline
\textbf{OuterSPACE} \cite{outerspace} & 2018 & Outer Product parallelized across rows of $A$ & Sparse matrix multiply with serial multiply/add phases, custom merge unit \\ \hline
\textbf{ExTensor} \cite{extensor} & 2019 & Inner Product tiled across all dimensions for locality & Arbitrary Einsums and TACO formats \cite{taco_formats}, skip-ahead intersection unit \\ \hline
\textbf{MatRaptor} \cite{matraptor} & 2020 & Row-wise Product with parallel summation & Sparse matrix multiply with co-design of micro-architecture and C$^2$SR format \\ \hline
\textbf{SIGMA} \cite{sigma} & 2020 & Inner Product parallelized across multiple dimensions & Sparse matrix multiply with custom bitmap format, flexible hardware topology \\ \hline
\textbf{SpArch} \cite{sparch} & 2020 & Outer Product with parallel merge & Sparse matrix multiply with optimized RAM interface in sum  phase \\ \hline
\textbf{Tensaurus} \cite{tensaurus} & 2020 & Inner Product with extended scalar-fiber product followed by fiber-fiber product ($SF^3$) &   $SF^3$ demonstrated  applicability to multiple Einsums beyond matrix-matrix multiply \\ \hline
\textbf{Gamma} \cite{gamma} & 2021 & Row-wise Product, adoption of Gustavson's alg. & Sparse matrix multiply with custom FiberCache, transposed merge-and-sum
\end{tabular}
\vspace{-0.1in}
\end{table*}

\subsection{Tensor Algebra with Extended Einsums}
\label{sec:background:einsum}

\specname expresses the individual computations performed by an accelerator using equations written in
an extended Einstein summation notation~\cite{einsum, sze:2020:epo, extensor}. 
For simplicity, we call equations written in this form Einsums.
Einsums are general enough to describe all tensor algebra kernels and have been used as the tensor algorithm specification in prior work for tensor algebra compilation~\cite{taco, unified_convolution_framework} and accelerator modeling~\cite{timeloop, sparseloop_micro, sze:2020:epo}.

We now present an operational definition of the Einsums used by \specname.
An Einsum specifies three things: (1) the input and output tensors involved and their ranks, (2) an \emph{iteration space} containing a point for each computation to be performed, and (3) the specific computation to be performed at each point in the iteration space.
For example, the Einsum for matrix-vector multiply is:
\begin{equation}
    Z_{m}= A_{m,k} \times B_{k}. \label{eqn:mv}
\end{equation}
Here, the equation defines the input tensors ($A$, $B$), the output tensor ($Z$), and the iteration space---the Cartesian product of all legal coordinates in the expression---($M \times K$). An implementation of this Einsum must traverse each point in this space. For each point, it computes the operation ($\times$) on the right-hand-side using the specified points in the input operands ($A, B$). It then takes the result and populates the location specified in the left-hand-side ($Z$).
Since the $K$ rank does not appear in the output tensor, the Einsum will attempt to repeatedly populate the same point ($Z_m$).
Einsum semantics resolve this by sequentially reducing the multiple values 
(using, in this case, addition) into a single value for that point.
Note that the Einsum does not specify iteration order; this is left to the mapping (Section~\ref{sec:background:densemapping}).

By adding a new rank $N$ to $B$, we can extend the above Einsum to matrix multiplication:
\begin{equation}
    Z_{m,n}= A_{m,k} \times B_{k,n}. \label{eqn:mm}
\end{equation}
\noindent This expands the iteration space to $M \times K \times N$. 

As another example, our operational definition of an Einsum allows us to represent kernels beyond standard tensor algebra.
For example, we can write 1D direct convolution with the Einsum:
\begin{equation}
O_{q} = I_{q+s} \times F_s \label{eqn:1dconv-direct}
\end{equation}
\noindent Just like the other examples, this equation defines the input tensors ($I$, $F$), output tensors ($O$), the iteration space ($Q \times S$), and the computation to occur at each point in the space.
A reduction occurs across points in the iteration space with the same $q$ and different $s$ coordinates, as one would expect from 1D convolution.

\subsection{Mapping Hardware Accelerators}
\label{sec:background:densemapping}

\emph{Mapping}~\cite{eyeriss} is the task of scheduling the computation of an Einsum onto limited hardware resources to jointly optimize for the desired combination of throughput, latency (execution time), power, etc.
We summarize the mapping attributes used for hardware modeling and design in prior work~\cite{interstellar, timeloop, maestro, cosa, mindmappings, tvm, zigzag} that we use throughout. 

\textbf{1) Loop order.}
The Einsum's large iteration space must be serialized through finite datapath resources in some order.
Two choices for Equation~\ref{eqn:mv} are: (1) 
$[M, K]$ or (2) $[K, M]$.
Loop order is read left-to-right, corresponding to the topmost-to-bottommost loop in a loop nest.
For example, (1) above reads ``for each value of $m$, iterate through all values of $k$.''
This choice affects data locality and, in turn, memory access costs.
Depending on on-chip buffer sizes, loop order (1) for matrix-vector multiply keeps an element of $Z$ \emph{stationary}~\cite{eyeriss} in on-chip memory while $B$ is streamed in multiple times.
Meanwhile, loop order (2) keeps $B$ stationary but repeatedly streams $Z$.

\textbf{2) Splitting.}
\label{sec:background:map_space:partitioning}
To further improve data locality across all tensors, many algorithms employ splitting (or strip mining, blocking, etc.) to divide the iteration space into subspaces that refer to a small enough subset of each of the tensors that they fit fully in on-chip buffers. 
Fibertrees model these subsets by partitioning their fibers according to the split iteration space.
How a fiber is partitioned is a function of the coordinates in the fiber.
For example, suppose matrix $A$ has a rank-order of $[M, K]$.
Splitting $K$ by shape $K0$ results in a new $A$ tensor with rank-order $[M, K1, K0]$,
where $K$ is split into $K1$ partitions with $K0$ coordinates each.\footnote{In other words, each tile stores coordinates in the coordinate range
$[i*K0, (i+1)*K0)$ for some $i$.}
    
\textbf{3) Work scheduling.}
\label{sec:background:map_space:spacetime}
Finally, the mapping specifies how the iteration space is traversed, by placing each point at a specific location in both time and space.
Mapping a computation at different locations in time implies that the computation is serial (i.e., computations happen one after another on the same component), while mapping at different locations in space implies parallelism (i.e., computation happens at the same time on different processing elements (PEs)).

\subsection{Accelerating Sparse Tensor Algebra}
\label{sec:background:sparse}

Sparse tensor algebra introduces new opportunities and challenges to the mapping problem.
Sparse tensors are typically \emph{compressed} to remove the zero elements, resulting in fibertrees with missing coordinate-payload pairs. 
Compression can yield significant savings in storage and data transfers and avoid ineffectual compute---operations that have no impact on the result and can be safely skipped, e.g., multiplication or addition with zero~\cite{extensor}.

However, realizing these benefits requires accelerators to ``sparsify'' the iteration space, or remove the ineffectual compute, increasing design complexity, sensitivity to memory latency/bandwidth, and load imbalance.
For example, the $[M, K]$ loop order for Equation~\ref{eqn:mv} may \emph{skip}, for example, from $(m=0,k=2)$ to $(m=2,k=0)$ in one step~\cite{sparseloop_micro}.
Such skipping can remove ineffectual compute but may require \emph{co-iteration} of the operands and additional operations (e.g., intersection for fibers multiplied together).
Without careful engineering, this can lead to 
inefficiencies that do not occur when tensors are dense~\cite{tactile}.
For example, the same-shape tiles produced by the scheme described in Section~\ref{sec:background:map_space:partitioning} may have different memory footprints, leading to data transfer and compute load imbalance when tiles are distributed to workers.

To deal with 
these challenges, sparse tensor accelerators have proposed a wide variety of 
custom hardware solutions, summarized in Table \ref{motivation:accelerators}. We note that the complexity of this topic makes such a table an imprecise and ultimately unsatisfying comparison.
Additionally, 
all of these works used custom hand-written simulators run on actual data sets 
to ensure all complexities are captured.
In the remainder of this paper we present a formalism to resolve this imprecision and enable concise apples-to-apples comparison. 
\section{Overview and Insights}
\label{sec:insights}

We now propose
\specname: a language and simulator generator
that 
1) enables the concise specification of a sparse tensor algebra accelerator and
2) generates efficiency statistics for that accelerator computing on actual sparse tensors.

Our key conceptual contribution, which guides the design of \specname, is to show that recent sparse tensor algebra accelerators can be expressed as \emph{cascades} (directed, acyclic graphs; DAGs) of mapped Einsums (Sections~\ref{sec:background:einsum}, \ref{sec:background:densemapping})
and \emph{content-preserving transformations} on their tensors.
This can be elaborated as two novel insights:

\smallskip
\noindent\textbf{Insight 1: Einsum cascades can represent multi-phase accelerator designs (Section~\ref{sec:insights:einsums}).}
A variety of accelerators and algorithms targeting seemingly monolithic kernels are more accurately and succinctly described as a sequence of distinct, interconnected phases.
We show that cascades of Einsums are sufficiently expressive to represent these multi-phase computations (e.g., Toeplitz-based convolution, OuterSPACE's multiply-merge, SIGMA's pre-filtering, Graphicionado's process and apply).

\smallskip
\noindent\textbf{Insight 2: Content-preserving transformations on tensors, representable as core operations on fibertrees, capture idioms for sparse tensor data orchestration (Section~\ref{sec:insights:transform}).}    
A variety of sparse accelerator behaviors (e.g., work scheduling, splitting, sorting/merging) can be represented as content-preserving transformations on \emph{specific} tensors in the Einsum cascade.  We show how these transformations can be represented as a small set of core operations performed on fibertrees (Section~\ref{sec:background:tensors}).
In specific, 
we use
fibertree rank partitioning/flattening as a general pattern for representing both sparse tensor splitting and work scheduling strategies.
Additionally, we use fibertree rank swizzling
as a general pattern for sorting and merging, which is often used to improve tensor traversal efficiency. 

\smallskip
This section describes the above insights in more detail
and how they enable the design of the \specname specification language.
Section~\ref{sec:model} describes how the \specname simulator generator converts \specname specifications into an imperative-style IR describing operations on fibertrees and how this IR (augmented with some additional information, e.g., describing the architecture and concrete formats) is subsequently converted into an accurate performance model.

\textbf{\specname specifications.}
The \specname \emph{specification language} is a declarative, domain-specific language (DSL) 
that defines the computation as a cascade of Einsums (\emph{expressions}), attributes on each tensor (\emph{declaration, rank-order, partitioning}), and a dataflow that describes when and where those tensors’ data  is moved
(\emph{loop-order, spacetime}).
We refer to the tensor declarations and Einsums as the \emph{einsum specification} and the tensor and dataflow attributes as the \emph{mapping specification}.
Rank swizzling is not expressed explicitly, but is inferred from other mapping attributes, such as the \emph{rank-order} and \emph{loop-order}.
\newcommand{\tech}[1]{$#1$}

\textbf{OuterSPACE running example.}
Throughout this section and Section~\ref{sec:model}, 
we use the example \specname specification in Figure~\ref{fig:outerspace:spec}, which describes the OuterSPACE accelerator~\cite{outerspace}.
At a high level, OuterSPACE accelerates SpMSpM
using the \emph{multiply-merge} algorithm.
It first performs all multiplications between input tensors \tech{A} and \tech{B} in an outer-product fashion, writes the resulting partial products to an array-of-linked-lists data structure, sorts the linked lists to facilitate reduction, and finally performs reductions over the now-sorted lists to derive final results. 
Throughout the rest of the section, we will discuss how \specname both implicitly and explicitly captures these behaviors.

\begin{figure}
\begin{lstlisting}[basicstyle=\small, commentstyle=\color{teal}]
einsum:
  declaration:   # Ranks are listed alphabetically in this section
    A: [K, M]     # Rank order is specified below in rank-order
    B: [K, N]
    T: [K, M, N]
    Z: [M, N]
  expressions:
    - T[k, m, n] = A[k, m] * B[k, n] # |\textcolor{teal}{$T_{k, m, n} = A_{k, m} \times B_{k, n}$}| |\label{teaal:einsum:t}|
    - Z[m, n] = T[k, m, n]  # |\textcolor{teal}{$Z_{m, n} = T_{k, m, n}$}| |\label{teaal:einsum:z}|
mapping:
  rank-order:
    A: [K, M] |\label{teaal:ro:a}|
    B: [K, N] |\label{teaal:ro:b}|
    T: [M, K, N] |\label{teaal:ro:t}|
    Z: [M, N] |\label{teaal:ro:z}|
  partitioning:
    T: |\label{teaal:part:t}|
      (K, M): [flatten()] |\label{teaal:part:t:flat}|
      KM: [uniform_occupancy(A.256), uniform_occupancy(A.16)] |\label{teaal:part:t:pe}|
    Z:
      M: [uniform_occupancy(T.128), uniform_occupancy(T.8)] |\label{teaal:part:z:pe}|
  loop-order:
    T: [KM2, KM1, KM0, N] |\label{teaal:lo:t}|
    Z: [M2, M1, M0, N, K] |\label{teaal:lo:z}|
  spacetime:
    T:
      space: [KM1, KM0] |\label{teaal:space:t}|
      time: [KM2, N]                  
    Z:
      space: [M1, M0] |\label{teaal:space:z}|
      time: [M2, N, K]
\end{lstlisting}
\vspace{-0.15in}
\caption{\small 
\specname specification for the Einsums and mappings of OuterSPACE~\cite{outerspace}, described in detail in Section~\ref{sec:insights}.
}
\label{fig:outerspace:spec}
\end{figure}

\subsection{Insight 1: Einsum cascades capture multi-phase accelerators}
\label{sec:insights:einsums}

\begin{table}[t!]
\footnotesize
\begin{minipage}{\columnwidth}
\caption{\label{table:cascades} \small Cascades of Einsums for various accelerators and algorithms.  Also see Figure~\ref{fig:novel:specs} for the Einsum cascades describing Graphicionado~\cite{graphicionado} and GraphDynS~\cite{graphdyns}.}
\vspace{-0.05in}
\centering
\footnotesize
\begin{tabular}{p{3.6cm}|p{3.9cm}}
\textbf{Accelerators / Algorithms} & \textbf{Cascade} \\ 
\hline \hline
ExTensor~\cite{extensor}'s SpMSpM & $Z_{m, n} = A_{k, m} \times B_{k, n}$ \\ \hline
Gamma~\cite{gamma}'s SpMSpM & $T_{k, m, n} = take(A_{k, m}, B_{k, n}, 1)$ \newline $Z_{m, n} = A_{k, m} \times T_{k, m, n}$ \\ \hline
OuterSPACE~\cite{outerspace}'s SpMSpM & $T_{k, m, n} = A_{k, m} \times B_{k, n}$ \newline $Z_{m, n} = T_{k, m, n}$ \\ \hline
SIGMA~\cite{sigma}'s SpMSpM & $S_{k, m} = take(A_{k, m}, B_{k, n}, 0)$ \newline $T_{k, m} = take(A_{k, m}, S_{k, m}, 0)$ \newline $Z_{m, n} = T_{k, m} \times B_{k, n}$ \\ \hline
Eyeriss~\cite{eyeriss}'s CONV & $O_{b, m, p, q} = I_{b, c, p + r, q + s} \times F_{c, m, r, s}$ \\ \hline
Toeplitz expansion/im2col + CONV~\cite{sze:2020:epo} & $T_{b, c, p, q, r, s} = I_{b, c, p + r, q + s}$ \newline $O_{b, m, p, q} = T_{b, c, p, q, r, s} \times F_{c, m, r, s}$ \\ \hline
Tensaurus~\cite{tensaurus}'s MTTKRP & $C_{i, r} = T_{i, j, k} \times B_{j, r} \times A_{k, r}$ \\ \hline
Factorized MTTRKP~\cite{mttkrp-fpga} & $S_{i, j, r} = T_{i, j, k} \times A_{k, r}$ \newline $C_{i, r} = S_{i, j, r} \times B_{j, r}$\\ \hline
Cooley-Tukey FFT Step~\cite{cooley-tukey-fft} & $E_{0, k0} = P_{0, k0, n1, 0} \times X_{n1, 0}$ \newline $O_{0, k0} = P_{0, k0, n1, 0} \times X_{n1, 1}$ \newline $T_{k0} = P_{0, k0, 0, 1} \times O_{0, k0}$ \newline $Y_{0, k0} = E_{0, k0} + T_{k0}$ \newline $Y_{1, k0} = E_{0, k0} - T_{k0}$ \\
\end{tabular}
\end{minipage}
\vspace{-0.1in}
\end{table}

Our first insight is that seemingly monolithic tensor algebra kernels (e.g., matrix multiply) are often implemented as a DAG of operations, and that each of these operations can be expressed as an Einsum that produces and consumes intermediate tensors.
We call this DAG a \emph{cascade}.
For example, consider a 1D convolution between input \tech{I} and filter \tech{F}.
Convolution is performed using two predominant implementation styles.
The first style is direct convolution, which is often employed by accelerators.  
As an Einsum, direct convolution is written as
\begin{equation}
O_{q} = I_{q+s} \times F_s \label{eqn:1dconv-direct}
\end{equation}
An alternative style is the Toeplitz expansion~\cite{sze:2020:epo}, which converts the convolution into a matrix-vector or matrix-matrix multiply and is common on systolic arrays and data-parallel processors like GPUs.
First, the input is refactored into a matrix to enable, in this case, matrix-vector  multiplication between the input (now stored in \tech{T}) and the filter \tech{F} in the second stage.
An important observation is that this can be written as the following sequence of dependent Einsums:
\begin{equation}
T_{q, s} = I_{q + s};\;\;\;\; O_{q} = T_{q, s} \times F_s
\label{eqn:1dconv-toeplitz}
\end{equation}
Importantly, the RHS of the Einsum used to generate \tech{T} mirrors how \tech{I} is indexed in the Einsum for direct convolution.
The Toeplitz expansion simply relaxes the requirement that the access into \tech{I} and the corresponding access into \tech{F} happen at the same time.
Decomposing an Einsum into a cascade enables each resulting Einsum to be mapped independently, exposing new degrees of freedom for building and using intermediate tensors.
Note that this sequence of Einsums says nothing about how (if at all) the two stages are overlapped.
They can happen entirely sequentially, or the accelerator can implement pipeline parallelism.
For example, the \tech{Q} rank can be partitioned (Section~\ref{sec:insights:partitioning}) and once a partition of \tech{T} is produced, it can be consumed by the multiply stage.
Section~\ref{sec:model:toolflow} describes how \specname determines this parallelism.

Beyond convolution, cascades of Einsums can be used to represent other common implementation styles in sparse tensor algebra accelerators.
For example, they capture the  multiply-merge algorithm in the OuterSPACE example (Figure~\ref{fig:outerspace:spec}).
During the multiply phase (Line~\ref{teaal:einsum:t}), columns of the \tech{A} matrix are multiplied with rows of the \tech{B} matrix to form partial products, which we call \tech{T}.
Then, during the merge phase, specified by the second Einsum (Line~\ref{teaal:einsum:z}), \tech{T} is reduced along the \tech{K} rank, yielding the final result \tech{Z}.

Sparsity also motivates new operations on tensors, and our Einsum notation can be extended to include them if needed.
We currently support one---the $\mathsf{take(.)}$ operator---which decouples intersection
from computation
with the following semantics: if at least one of the inputs is zero at a point, the output is zero, otherwise, copy one of the inputs into the output. 
Take the example:
\begin{equation}
T_{k, m, n} = take(A_{k, m}, B_{k, n}, 1)
\end{equation}
The final parameter denotes which input is copied into the output.
This example copies $B$ into $T$, but if the last parameter were $0$, $A$ would be copied.

Beyond the above examples, Table~\ref{table:cascades} shows a variety of accelerators and algorithms represented as cascades of Einsums.

\begin{figure*}
  \centering
      \includegraphics[width=0.9\textwidth]{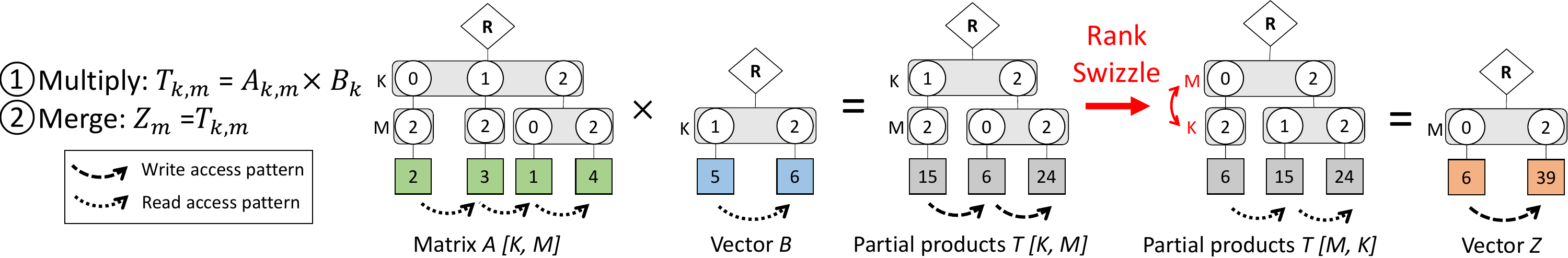}
      \vspace{-0.05in}
  \caption{\label{fig:insights:swizzle} \small 
            Rank swizzling
            in sparse tensor algebra computations, using outer-product multiply-merge matrix-vector multiplication.
            Matrix $A$ and vector $B$ use values from Figure~\ref{fig:background:fibertree} for consistency.
            An offline rank swap ensures that $A$ has rank order $[K, M]$ prior to the multiply phase,
            and an online rank swap ensures that $T$ has rank order $[M, K]$ prior to the merge phase,
            ensuring concordant traversal in both phases.
            }
\vspace{-0.15in}
\end{figure*}

\subsection{Insight 2: Content-preserving transformations on fibertrees capture accelerator data-orchestration strategies}
\label{sec:insights:transform}

Our second insight is that a variety of accelerator behaviors (e.g., work scheduling) are describable as what we call \emph{content-preserving transformations} applied to \emph{specific} tensors in the Einsum cascade.
We say a tensor transformation is content-preserving if it does not change the content of the tensor, i.e., the set of values at the leaves of the fibertree, but does change 
the coordinate system used to access each value. 
Such transformations may also impact the tensor's data layout when lowered to a concrete representation.

We make an important observation that the core operations performed on fibertrees (Section~\ref{sec:background:tensors}) represent a set of content-preserving transformations that are useful for describing a variety of prevalent sparse data orchestration strategies.
Specifically, fibertree rank partitioning (and its inverse: flattening) can be used as a single abstraction for specifying both 
sparse tensor splitting and work scheduling strategies.
Similarly, fibertree rank swizzling can be used as a single abstraction for specifying transposing data in memory, sorting, and merging.

\subsubsection{Sparse Tensor Splitting and Work Scheduling}
\label{sec:insights:partitioning}

Recall from Section~\ref{sec:background:map_space:partitioning} that splitting for dense problems is shape-based. 
This can be expressed by partitioning a fibertree rank at coordinate-based boundaries given by the tile dimension.
Unfortunately, when data is sparse, this strategy
can lead to low reuse and under-utilization of tiles (and therefore buffers)~\cite{tactile}, i.e., if different partitions have different occupancies.

We make an important observation that partitioning naturally generalizes to other types of splitting that \emph{can} adapt to irregular sparsity, simply by changing the \emph{partitioning criteria}, i.e., where the partition boundaries occur. 
From studying existing accelerators, we define a simple sparsity-aware strategy we call \emph{uniform occupancy-based partitioning}.
In this scheme, each fiber at a level in the fibertree is split so that each new fiber has an equal number of elements (modulo remainders).
Importantly, each fiber's coordinate range 
after an occupancy-based partitioning is irregular.
Thus, to ensure that partitions of multiple tensors have matching coordinate ranges for co-iteration (Section~\ref{sec:background:sparse}), occupancy-based partitioning uses a \emph{leader-follower} paradigm: the partitions' coordinate ranges are chosen so that the leader tensor's partitions are equal occupancy and all follower tensors adopt those 
ranges~\cite{sparseloop_micro}.

Unfortunately, uniform occupancy-based partitioning may still result in partitions with varying occupancies because a partition must end where its parent fiber ends.
Flattening (Section~\ref{sec:background:tensors}), when combined with occupancy-based partitioning, mitigates this imbalance by first combining the flattened ranks, then redistributing the elements so that, globally, each partition has the same number of values.
For example, Figure~\ref{fig:insights:partitioning} shows how a fibertree whose fibers start with an unequal number of coordinates can be flattened and re-partitioned to equalize the number of coordinates per partition.
Note that, though all partitioning directives modify the fibertree abstract representation, the concrete representation may remain unchanged.

The above describes how tensor data can be efficiently split in the presence of sparsity.
More subtly, we observe that partitioning and flattening are also useful abstractions through which to specify \emph{work scheduling} when work is parallelized.
Consider OuterSPACE, which works on 256 non-empty elements of matrix \tech{A} at a time during the multiply phase, further subdividing these into 16 groups of 16 elements to each be processed by a ``Processing Tile'' (a group of PEs~\cite{outerspace}).
The \specname specification for OuterSPACE (Figure~\ref{fig:outerspace:spec}) represents this as a flattening (Line~\ref{teaal:part:t:flat}) and then an occupancy-based partitioning applied twice hierarchically (Line~\ref{teaal:part:t:pe}).\footnote{
Note: OuterSPACE only enables half its PEs during the merge step, so the occupancy-based partitioning applied to the second Einsum (Line~\ref{teaal:part:z:pe}) only involves 128 PEs (8 per ``Processing Tile'').}
The \specname specification describes the \emph{parallelism} this partitioning enables on Line~\ref{teaal:space:t} by scheduling ranks \tech{KM1} and \tech{KM0} in space (Section~\ref{sec:background:map_space:spacetime}).

\subsubsection{Transposition, Sorting, and Merging}
\label{sec:insights:transform:swizzling}

We observe that sparse tensor algebra accelerators employ a number of techniques that, when expressing their tensors in the fibertree abstraction, are tantamount to fibertree rank swizzles.
These operations enable the more efficient
\emph{concordant} (as opposed to \emph{discordant}) traversal~\cite{sze:2020:epo}.
Concordant traversal occurs when a loop nest traverses a fibertree in the order in which its ranks appear, i.e., traverses each fiber sequentially and in a depth-first manner.
For example, in Figure~\ref{fig:insights:swizzle}, \tech{A} is traversed concordantly, since it has a \tech{[K, M]} rank order, and the multiply phase has a \tech{[K, M]} loop order.
Thus, we never have to search for the next $k$ or $m$ coordinate; it is always the first or next coordinate in the current fiber.\footnote{Though many concrete representations enable efficient sequential iteration through fibers, some do not.
The true cost of iteration is accounted for during modeling (Section~\ref{sec:model}).}
Conversely, iterating over \tech{K} in the bottom-most loop would be a discordant traversal (for this rank order).
In OuterSPACE (Figure~\ref{fig:outerspace:spec}), despite all of the partitioning, during the multiply phase both \tech{A} and \tech{B} are traversed concordantly.
\tech{K} is traversed sequentially and most slowly, then \tech{M} and then \tech{N}.

It is common practice to swizzle ranks to enable concordant traversal on input tensors.
For example, the transposition of a matrix from the CSR format into the CSC format can be viewed as a rank swizzle
and is used by OuterSPACE to achieve a \tech{[K, M]} rank order on \tech{A} (Line~\ref{teaal:ro:a}) in preparation for the outer-product-style multiply phase. 
Input tensor swizzles are usually performed offline.

More subtly, we observe that sparse tensor accelerators also perform rank swizzles on \emph{intermediate tensors} formed during kernels expressed as cascades of Einsums
(Section~\ref{sec:insights:einsums}).
Depending on whether coordinates in the intermediate tensors are stored in sorted or unsorted order and on the extent to which the intermediate tensors are built before being consumed, this either requires a merge or a (more expensive) sort operation.

Figure~\ref{fig:insights:swizzle} shows an example of a multiply-merge for outer product, matrix-vector multiplication.
To support concordant traversal on both input and output tensors, the multiply phase uses a \tech{[K, M]} loop order, while the merge phase uses an \tech{[M, K]} loop order.
Thus, at the end of the multiply phase, \tech{T} has rank order \tech{[K, M]}.
Then, during the reduction, a rank swizzle changes the rank order of $T$ to \tech{[M, K]} to match the \tech{[M, K]} loop order.
The dashed arrows in Figure~\ref{fig:insights:swizzle} show that \emph{both} the tensor read and write access patterns through $A$, $B$, $T$, and $Z$ are all concordant through both phases.
Importantly, such online rank swizzles 
may
degrade performance and, therefore, warrant dedicated hardware support.
Yet, they can significantly improve spatial/temporal locality, and thus, appear in multiple prior designs~\cite{gamma, outerspace, sparch}.

By default, \specname infers rank swizzling automatically to maintain concordant traversal.
For example, for OuterSPACE (Figure~\ref{fig:outerspace:spec}), \tech{T} has rank order \tech{[M, K, N]}, but \specname produces \tech{T} during the multiply phase in \tech{[K, M, N]} order, swizzles it to \tech{[M, K, N]} order to be stored in memory, and then swizzles it again to \tech{[M, N, K]} order to prepare for the merge.

\section{Generating the Model}
\label{sec:model}

\begin{figure*}[ht]
\centering
\includegraphics[width=0.95\textwidth]{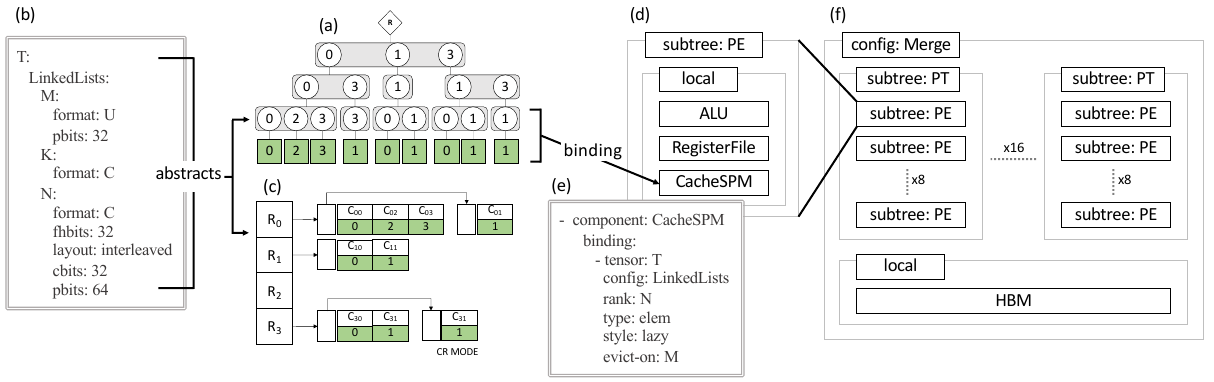}
\vspace{-0.05in}
\caption{\specname concrete/hardware-level model of the OuterSPACE accelerator~\cite{outerspace}.
The fibertree (a) combined with the format specification (b) describe the concrete representation, a custom array-of-linked-lists format (c).
\specname specifies the architecture hierarchically (f), where each level has a set of local components (d) that have tensor operations bound to them (e).
More details are given in Section~\ref{sec:model:outerspace}.
}
\label{fig:model:outerspace}
\end{figure*}

\begin{figure}[ht]
\centering
\includegraphics[width=0.8 \columnwidth]{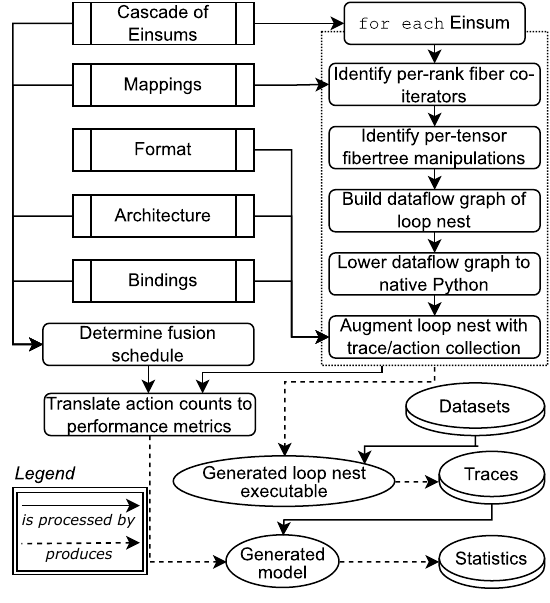}
\vspace{-0.05in}
\caption{\specname tool flow diagram, described in detail in Section~\ref{sec:model:toolflow}.
}
\vspace{-0.05in}
\label{fig:model:toolflow}
\vspace{-0.25in}
\end{figure}

\begin{figure}[ht]
\centering
\includegraphics[width=\columnwidth]{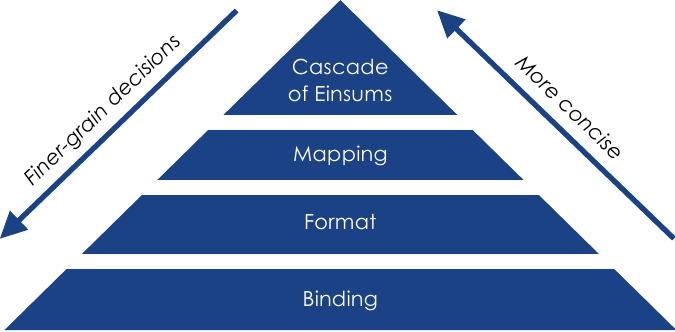}
\caption{Separation of concerns enabled by \specname.
We consider the architecture to be fixed during map space exploration and, therefore, outside the separation of concerns.
}
\label{fig:model:pyramid}
\vspace{-0.15in}
\end{figure}

In Section~\ref{sec:insights}, we showed that the fibertree abstraction is general enough to describe many of the design
decisions used in sparse tensor algebra accelerators.
However, 
to manifest a specific design, the fibertrees must be lowered onto concrete representations and their operations bound to specific hardware components.
In this section, we define three additional specifications---\emph{format}, \emph{architecture}, and \emph{binding}---used by \specname to perform this lowering and describe how these, plus the \emph{einsum} and \emph{mapping} specifications from Section~\ref{sec:insights}, are combined to produce
an executable model for evaluating
accelerator workload performance.

\subsection{Lowering Mapped Einsums to Hardware}
\label{sec:model:specs}

\begin{table}[t!]
\footnotesize
\begin{minipage}{\columnwidth}
\caption{\label{spec:hw-specs} \small Supported hardware components and their attributes.}
\vspace{-0.1in}
\centering
\begin{tabular}{p{1.3cm}|p{6.4cm}}
\textbf{Component} & \textbf{Attributes} \\ \hline \hline
DRAM & bandwidth \\ \hline
Buffer & type (buffet~\cite{buffets} or cache), width, depth, bandwidth \\ \hline
Intersection & type (two-finger, leader-follower, or skip-ahead), leader \\ \hline
Merger & inputs, comparator\_radix, outputs, order (fifo, opt), reduce \\ \hline
Sequencer & num\_ranks \\ \hline
Compute & type (mul or add) \\
\end{tabular}
\end{minipage}
\vspace{-0.2in}
\end{table}

This section describes the three additional specifications (\emph{format}, \emph{architecture}, and \emph{binding}) that are used to lower mapped Einsums to concrete representations and hardware resources.

\subsubsection{Format}
\label{sec:model:specs:format}

Prior works on modeling sparse tensor algebra computations~\cite{sparseloop_micro,stonne} extend TACO's level formats concept~\cite{taco_formats}, which facilitates a convenient abstraction for translating a fibertree to its concrete representation
by specifying a per-fiber format as described in~\cite{sze:2020:epo}.
However, the formats used by existing sparse accelerator modeling frameworks restrict themselves to a fixed number of common configurations (e.g., bitmap~\cite{stonne} or uncompressed offset pointers~\cite{sparseloop_micro}).

\specname extends these concepts with a more modular specification, capturing a larger class of formats than existing tools, by separating the attributes of a fiber's format into three categories: a format type, a layout, and data widths for the coordinates (\tech{cbits}), payloads (\tech{pbits}), and fiber headers (\tech{fhbits}).
Fibers are concretized as an array of coordinates and a array of payloads (struct-of-arrays) or as a single array of elements (array-of-structs).
\specname supports three format types: uncompressed (\tech{U})---where the sizes of the
data arrays correspond to the shape of the fiber, compressed (\tech{C})---where the sizes of the data arrays correspond to the occupancy of the fiber, or a combination (\tech{B})---where the coordinates are uncompressed and the payloads are compressed.
For simplicity, currently,
all fibers in a rank have the same format.
Note that not all information in the fibertree is stored explicitly, e.g., an uncompressed fiber does not need to store coordinates because they can be inferred from the position of the payloads.
\specname supports this by allowing the corresponding data width---in this case,
\tech{cbits}---to be unspecified or set to 0.

This specification is flexible enough to support a variety of common formats (e.g., uncompressed arrays, CSR, run-length encoding) and custom formats
even beyond those supported by TACO (e.g., from OuterSPACE~\cite{outerspace} and SIGMA~\cite{sigma}).
The specification also allows each tensor to be associated with multiple formats (differentiated by the configuration name), since manipulating the fibertree may cause the representation to change dynamically.
Finally, while there are formats that require functionality outside the above framework (e.g., TACO's hashed format, which requires a hash function be specified), we feel these could be added with minimal changes to the both other specifications and simulator generator.

\subsubsection{Architecture}
\label{sec:model:specs:arch}

The \specname architecture specification (inspired by Time\-loop~\cite{timeloop}) describes the accelerator topology as a tree of compute and storage units.
At each level of the hierarchy, one can define a list of hardware components local to that level and a list of subtrees below that level.
In addition to the component classes supported by Timeloop, we define new classes of components that are involved in performance-limiting operations on sparse accelerators, including caches, intersection units, and hardware mergers.
Table~\ref{spec:hw-specs} gives the full list of supported classes and their attributes.
Since an accelerator (e.g., OuterSPACE~\cite{outerspace}) may reorganize itself during the execution of a single 
computation, \specname also supports specifying multiple topologies for the same design.

\subsubsection{Binding}
\label{sec:model:specs:binding}

Finally, \specname's binding specification matches the Einsum- and mapping-induced fibertree operations to specific concrete representations and hardware components in the architecture.
First, each Einsum must be bound to a single accelerator topology.
Then, for each storage component, its bindings describe what data resides there.
Each binding contains the data's tensor, configuration,
rank, type (e.g., payload),
whether elements are accessed lazily (loading/storing only the element on access) or eagerly (loading/storing the entire subtree below an element on access),
and sometimes (e.g., for buffets~\cite{buffets}), how long the data is buffered.
A storage component can have multiple such bindings.
Similarly, for each compute component, the binding describes which compute operations are performed on that component.

\subsubsection{\specname's Expressibility and Extensibility}
\label{sec:model:express}

Putting everything together: 
\specname decomposes the design of an accelerator into a set of categories of abstractions, where a specific design choice is an instance of the category.
As shown in Figure~\ref{fig:model:pyramid}, these abstractions are hierarchical; the accelerator's cascade of Einsums is the most concise representation of that accelerator's design, while its binding encodes the finest-grain design decisions enabling the highest-fidelity modeling.

This separation of concerns enables \specname to express a large number of accelerators, and facilitates the process of adding new features to represent yet other accelerators.
For example, \specname can express accelerators that differ only in specific details (such as tensor format~\cite{taco} or cache replacement policy~\cite{gamma}) simply by changing that part of the specification, leaving all else equal.
New features can likewise be added by augmenting the relevant abstraction category, again leaving other categories unmodified.

\subsection{Specifying OuterSPACE~\cite{outerspace}}
\label{sec:model:outerspace}

We continue to use OuterSPACE as a running example to motivate the features provided by \specname.
Figure \ref{fig:model:outerspace} shows a simplified version of OuterSPACE's custom tensor format, its architecture during the merge phase, and the correspondence between the fibertree representation of $T$ and its concrete representation and binding.
In Figure~\ref{fig:model:outerspace}a, we see the fibertree for a concrete example tensor $T$.
The format specification (Figure~\ref{fig:model:outerspace}b) for this tensor lowers it onto OuterSPACE's custom array-of-linked-lists format (Figure~\ref{fig:model:outerspace}c).
To differentiate it from other representations of the same tensor, we give it the configuration name \tech{LinkedLists}.
On the $M$ rank, the array of pointers is given by an uncompressed (\tech{U}) array of payloads.
On the $N$ rank, the fiber header data width (\tech{fhbits}) describes the linked list pointers, the \tech{layout} describes that corresponding coordinates and payloads are 
adjacent (array-of-structs), and the \tech{cbits} and \tech{pbits} describe the data widths of the coordinates and payloads, respectively.

Figure~\ref{fig:model:outerspace}d shows an OuterSPACE PE.
During the merge phase, this level has three
components: the ALU, the register file, and the L0 scratchpad.
OuterSPACE loads the entire subtree under a given $M$ coordinate into the L0 scratchpad 
to perform its sort.
\specname expresses this binding with the specification given in Figure~\ref{fig:model:outerspace}e.
The \tech{tensor}, \tech{config}, \tech{rank}, and \tech{type} denote exactly what data is buffered, and
the \tech{evict-on} keyword is required for binding to explicitly managed buffers, whose fill and drain policy must be set by the user.
Because the elements bound to this buffer evict on $M$, old data is drained when the $M$ coordinate changes.
Finally, Figure~\ref{fig:model:outerspace}f shows an overview of the entire accelerator topology.

\newsavebox{\gammas}
\begin{lrbox}{\gammas}
\begin{lstlisting}[basicstyle=\small]
einsum:
  declaration:
    A: [K, M]
    B: [K, N]
    T: [K, M, N]
    Z: [M, N]
  expressions:
    - T[k,m,n] = take(A[k,m], B[k,n], 1)
    - Z[m,n] = T[k,m,n] * A[k,m]
mapping:
  rank-order:
    A: [M, K]
    B: [K, N]
    T: [M, K, N]
    Z: [M, N]
  partitioning:
    T:
      M: [uniform_occupancy(A.32)]
      K: [uniform_occupancy(A.64)]
    Z:
      M: [uniform_occupancy(A.32)]
      K: [uniform_occupancy(A.64)]
  loop-order:
    T: [M1, M0, K1, K0, N]       
    Z: [M1, M0, K1, N, K0]
  spacetime:
    T:
      space: [M0, K1]                       
      time: [M1, K0, N]
    Z:
      space: [M0, K1]
      time: [M1, N, K0]
\end{lstlisting}
\end{lrbox}
\newsavebox{\extensor}
\begin{lrbox}{\extensor}
\begin{lstlisting}[basicstyle=\small]
einsum:
  declaration:
    A: [K, M]
    B: [K, N]
    Z: [M, N]
  expressions:
    - Z[m,n] = A[k,m] * B[k,n]
mapping:
  rank-order:
    A: [K, M]
    B: [K, N]
    Z: [M, N]
  partitioning:
    Z: |\label{line:e:part:start}|
      K:
        - uniform_shape(K1)
        - uniform_shape(K0)
      M:
        - uniform_shape(M1)                 
        - uniform_shape(M0)
      N:
        - uniform_shape(N1)
        - uniform_shape(N0) |\label{line:e:part:end}|
  loop-order:
    Z: [N2, K2, M2, M1, N1, K1, M0, N0, K0]
  spacetime:
    Z:
      space: [K1]
      time: [N2, K2, M2, M1, N1, M0, N0, K0]
\end{lstlisting}
\end{lrbox}
\begin{lrbox}{\sigma}
\begin{lstlisting}[basicstyle=\small]
einsum:
  declaration:
    A: [K, M]
    B: [K, N]
    S: [K, M]
    T: [K, M]
    Z: [M, N]
  expressions:
    - S[k, m] = take(A[k, m], B[k, n], 0) |\label{line:s:e1}|
    - T[k, m] = take(A[k, m], S[k, m], 0) |\label{line:s:e2}|
    - Z[m, n] = T[k, m] * B[k, n]   |\label{line:s:e3}|
mapping:
  rank-order:
    A: [K, M]
    B: [K, N]
    S: [K, M]
    T: [K, M]
    Z: [M, N]
  partitioning:
    Z:
      K: [uniform_shape(128)]         |\label{line:s:p:k}|
      (M, K0): [flatten()]              |\label{line:s:p:f}|
      MK0: [uniform_occupancy(T.16384)] |\label{line:s:p:mk0}|
  loop-order:
    S: [K, M, N]
    T: [K, M]
    Z: [K1, MK01, MK00, N]
  spacetime:
    S:
      space: []
      time: [K, M, N]
    T:
      space: []
      time: [K, M]
    Z:
      space: [MK00]                     |\label{line:s:space}|
      time: [K1, MK01, N.coord]
\end{lstlisting}
\end{lrbox}

\begin{figure*}[ht!]
  \centering
  \begin{subfigure}{.3\linewidth}
    \usebox{\gammas}
    \caption{\label{spec:gamma} \small Gamma accelerator~\cite{gamma}.}
  \end{subfigure} \hfill
  \begin{subfigure}{.3\linewidth}
    \usebox{\extensor}
    \caption{\label{spec:extensor} \small ExTensor accelerator~\cite{extensor}.}
  \end{subfigure} \hfill
  \begin{subfigure}{.3\linewidth}
    \usebox{\sigma}
    \caption{\label{spec:sigma} \small SIGMA accelerator~\cite{sigma}.}
  \end{subfigure} \hfill
  \vspace{-0.05in}
  \caption{\label{spec:accelerators}
  \small State-of-the-art sparse tensor accelerators.
  $\mathrm{uniform\_shape()}$/$\mathrm{flatten()}$ are syntax
  for shape-based partitioning/flattening (Section~\ref{sec:insights:partitioning}).
  }
  \vspace{-0.15in}
\end{figure*}

\subsection{Simulator Generation}
\label{sec:model:toolflow}

Figure~\ref{fig:model:toolflow} demonstrates how \specname puts everything together.
For each Einsum in the cascade, \specname combines the
Einsum equation with its mapping information to produce an executable loop nest.
To do so, it identifies the necessary per-tensor fibertree manipulations (e.g., rank swizzling) and per-rank fiber co-iterators (e.g., intersection).
\specname then uses this information to build a dataflow graph of a 
loop nest, which it then lowers to an embedded DSL within Python for executing computations as fibertree operations~\cite{hifiber}.
The resulting code is an imperative-style representation of the Einsum cascade (i.e., a series of loop nests, one per Einsum), which can directly evaluate real tensors represented as fibertrees.
\specname then breaks the modeling of an accelerator into three stages: generate \emph{traces} describing when each coordinate and each payload is accessed, calculate the action counts for each component from the traces, and combine the action counts from all components to produce summary statistics like execution time and energy.

\textbf{Trace generation.}
\specname combines information from the format, architecture, and binding to 
identify which traces need to be collected in preparation for performance modeling. 
It then instruments the mapped loop nests to collect the desired traces.
When executed, the mapped loop nests
perform the computation on fibertrees (storing real tensor data)
and generate a trace of when each coordinate and each payload is accessed.
Therefore, unlike an analytical model, \specname is able to fully capture the impact of each real tensor's specific sparsity patterns on the kernel's performance, significantly improving \specname's fidelity over that of analytical models.
We quantitatively explore this phenomenon in Section~\ref{sec:eval:validation} and Figure~\ref{fig:eval:time:extensor}.

\textbf{Trace consumption.}
\specname provides a library of per-component action count models (see Table~\ref{spec:hw-specs} for a full list).
It inserts calls to these component models after the loop nest, passing information about their specific attributes (e.g., buffer width and depth) and a list of traces to be read.
During the evaluation of the model, each component uses the traces generated to produce the action counts it performed.

\textbf{Action count consumption.}
\specname uses Accelergy~\cite{accelergy} to translate action counts to energy use and a custom analytical modeling/bottleneck analysis to translate the action counts to execution time. 
To compute execution time, \specname must first determine the Einsum \emph{blocks}, or sets of Einsums that are fused together.
Fusion occurs when Einsums communicate by sharing sub-tensors with each other (instead of entire tensors).
The full cascade of Einsums, mappings, architectures, and bindings are used to determine the Einsum blocks.
Specifically, \specname infers that Einsums can be fused together when three conditions are met:\footnote{These conditions for inferring fusion are not fundamental and can be changed if needed.}
\begin{itemize}
\item The Einsums use the same accelerator configuration.
\item The temporal ranks in all loop orders before the first spatial rank are the same.
\item Disjoint subsets of the non-storage components are each exclusively used by only one Einsum.
\end{itemize}

As a simple heuristic, \specname starts at the first Einsum and greedily fuses the successive Einsums together into a single block, until it cannot do so any more.
At which point, it starts a new block.
\specname sums together the action counts for each component performed by each block and then computes per-block, per-component execution times.
It then applies a bottleneck analysis: the execution time of the block is the execution time of the longest component, and the execution time of the cascade is the sum of the execution times of all of the blocks.
\section{Accelerator Specification}
\label{sec:accel}

Sections~\ref{sec:insights}-\ref{sec:model} used OuterSPACE \cite{outerspace} as a running example. 
We now describe the Einsums and mapping specifications for three other, state-of-the-art accelerators relevant to our evaluation, shown in Figure~\ref{spec:accelerators}.
We have modeled other accelerators that we omit for space, including Graphicionado~\cite{graphicionado} and GraphDynS~\cite{graphdyns} (Section~\ref{sec:eval:design}), Eyeriss~\cite{eyeriss}, Tensaurus~\cite{tensaurus}, Flexagon~\cite{flexagon}, and DSTC~\cite{dstc}.
We also omit the format, architecture, and binding specifications for brevity.

\textbf{Gamma~\cite{gamma}.}
Gamma (Figure~\ref{spec:gamma}) is a row-wise-style accelerator that uses a tightly-pipelined multiply-merge-style architecture to reduce partial output traffic and enable concordant traversal across both input and output tensors.
In Gamma's dataflow, a row of $A$ is combined and reduced with rows of $B$.
Gamma distributes rows of $A$ to each PE and, based on which values in each row are non-zero, the PE fetches a subset of the rows of $B$.
This filtering is implemented using the $\mathsf{take(.)}$ operator (Section~\ref{sec:insights:einsums}).
After being fetched to each PE, the rows of $B$ (which initially have rank order $[K, N]$) are sorted with hardware mergers to facilitate reduction over $K$.
Similar to OuterSPACE, this is expressed as a rank swizzle:
$T$ has rank order $[M, K, N]$ and the rightmost (bottommost) rank in the loop order for the Einsum computing $Z$ is $K$.
Hence, \specname inserts a rank swizzle on $T$, making its rank order $[M, N, K]$ in the context of the second Einsum.
Unlike OuterSPACE, the two Einsums in the cascade are fused together, per the criteria described in Section~\ref{sec:model:toolflow}.

\textbf{ExTensor~\cite{extensor}.}
ExTensor (Figure~\ref{spec:extensor}) employs a hybrid dataflow that is inner product-style at the innermost level.
ExTensor's two salient characteristics are the use of uniform shape-based partitioning (Section~\ref{sec:background:map_space:partitioning}) and hierarchical intersection.
Lines~\ref{line:e:part:start}-\ref{line:e:part:end} describe this partitioning, while hierarchical intersection is accounted for implicitly due to fibertree semantics (Section~\ref{sec:background:sparse}).
Note that our ExTensor specification includes details beyond the original paper
from private correspondence with the authors about the actual design of the simulator used for evaluation.

\textbf{SIGMA~\cite{sigma}.}
SIGMA (Figure~\ref{spec:sigma}) is a 
deep-learning accelerator that uses occupancy-based partitioning to
only distribute non-zero elements of the stationary matrix to PEs, reducing ineffectual compute.
While SIGMA can be configured to support $A$ and $B$-stationary dataflows, we only describe/evaluate the $A$-stationary dataflow here.
SIGMA utilizes an Einsum cascade (Section~\ref{sec:insights:einsums}),
first identifying empty $K$-fibers (rows) of $B$ (Line~\ref{line:s:e1}), removing them from $A$ (Line~\ref{line:s:e2}), and then performing the multiplication (Line~\ref{line:s:e3}).
We express the partitioning on Lines~\ref{line:s:p:k}-\ref{line:s:p:mk0} using a combination of shape-based partitioning, flattening, and occupancy-based partitioning (Section~\ref{sec:insights:partitioning}).
Finally, because all PEs work in parallel, the spatial dimension is $MK00$ (Line \ref{line:s:space}).
\section{Experimental Setup}
\label{sec:eval:method}

\begin{table}[t]
\footnotesize
\begin{minipage}{\columnwidth}
\caption{\label{spec:tensors}\small  Tensor data set characteristics.
The top 5 tensors are used in our validation study (Section~\ref{sec:eval:validation}); the bottom 3 in our new design study (Section~\ref{sec:eval:design}).}
\vspace{-0.1in}
\centering
\begin{tabular}{p{2.6cm}|p{1.7cm}|p{0.6cm}|p{1.8cm}}
\textbf{Matrix} & \textbf{Shape} & \textbf{NNZ} & \textbf{Domain} \\ \hline
\hline
wiki-Vote ($\mathsf{wi}$) & $8.3K \times 8.3K$ & 104K & elections \\ \hline
p2p-Gnutella31 ($\mathsf{p2}$) & $63K \times 63K$ & 148K & file-sharing \\ \hline
ca-CondMat ($\mathsf{ca}$) & $23K\times 23K$ & 187K & collab. net.\\ \hline
poisson3Da ($\mathsf{po}$) & $14K \times 23K$ & 353K & fluid dynamics \\ \hline
email-Enron ($\mathsf{em}$) & $37K \times 37K$ & 368K & email comms. \\ \hline
\hline
flickr ($\mathsf{fl}$) & $0.82M\times 0.82M$ & 9.8M & site crawl graph \\ \hline
wikipedia-20070206 ($\mathsf{wk}$) & $3.6M \times 3.6M$ & 42M & site link graph \\ \hline
soc-LiveJournal1 ($\mathsf{lj}$) & $4.8M \times 4.8M$ & 69M & follower graph \\ 
\end{tabular}
\end{minipage}
\vspace{-0.15in}
\end{table}

\begin{table}
\footnotesize
\begin{minipage}{\columnwidth}
\caption{\label{spec:accel-hw} \small Hardware configs, chosen to match original publications.}
\vspace{-0.1in}
\centering
\begin{tabular}{p{2cm}|p{5.6cm}}
ExTensor \cite{extensor} & 
1 GHz clock speed, 128 PEs,
64 kB PE buffer per PE, 30 MB LLC,
68.256 GB/s memory bandwidth
\\ \hline
Gamma \cite{gamma} & 
1 GHz clock speed, 64-way merger per PE, 32 PEs,
3 MB FiberCache,
16 64-bit HBM channels, 8 GB/s/channel \\ \hline
OuterSPACE \cite{outerspace} &
1.5 GHz clock speed, 16 PEs per PT, 16 PTs,
16 kB L0 cache per PT, 4 kB L1 cache per 4 PTs,
16 64-bit HBM channels, 8000 MB/s/channel \\ \hline
SIGMA \cite{sigma} & 
500 MHz clock speed, 128 PEs per FlexDPE, 128 FlexDPEs,
32 MB Data SRAM, 4 MB Bitmap SRAM,
960 GB/s SRAM bandwidth, 1024 GB/s HBM bandwidth
\\ \hline
Graphicionado \cite{graphicionado} & 
1 GHz clock speed, 8 streams,
64MB eDRAM,
68 GB/s memory bandwidth
\\
\end{tabular}
\end{minipage}
\vspace{-0.2in}
\end{table}

This section describes the details of the experimental set-up used in Sections~\ref{sec:eval:validation}-\ref{sec:eval:design} to evaluate the performance characteristics of concrete accelerators.

\textbf{Tensors.} To evaluate the \specname models,
we execute the models for the accelerators on a combination of randomly generated matrices with uniform sparsity and a set of matrices from SuiteSparse~\cite{florida} and SNAP~\cite{snapnets}, described in Table~\ref{spec:tensors}.

\textbf{Simulation Framework.}
We implement the accelerators by writing \specname specifications for their Einsums, mappings, formats, architectures, and bindings.
For each accelerator, we use the hardware parameters given in Table~\ref{spec:accel-hw}.
\specname uses Accelergy~\cite{accelergy} as a power model to convert the per-component action counts to an energy characterization.

\textbf{Baselines.}
To validate our results, we 
normalize our performance estimates using the same baseline as the original papers that published the relevant accelerators.
All accelerators' ``reported'' statistics  come either from published results or from private communication with the original authors.
When possible, we also report Sparseloop~\cite{sparseloop_micro} performance estimates using the \emph{hypergeometric} sparsity distribution on both the inputs and outputs, estimated using the values in Table~\ref{spec:tensors}, and the hardware parameters in Table~\ref{spec:accel-hw}.
\section{Simulator Validation}
\label{sec:eval:validation}

\begin{figure*}
\begin{subfigure}{.3\textwidth}
  \centering
  \includegraphics[width=\linewidth]{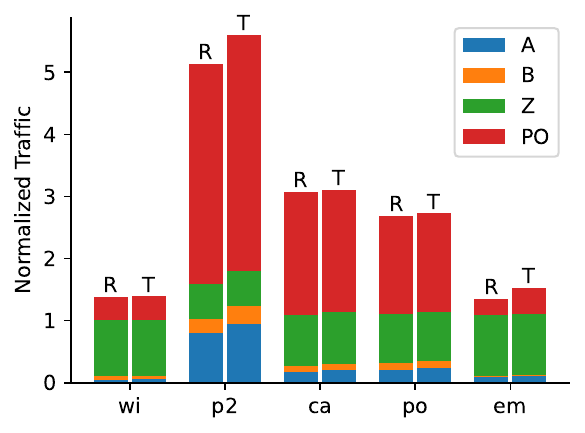}
   \vspace{-0.2in}
 \caption{ExTensor Memory Traffic}
  \label{fig:eval:mem:extensor}
\end{subfigure}
\hfill
\begin{subfigure}{.3\textwidth}
  \centering
    \includegraphics[width=\linewidth]{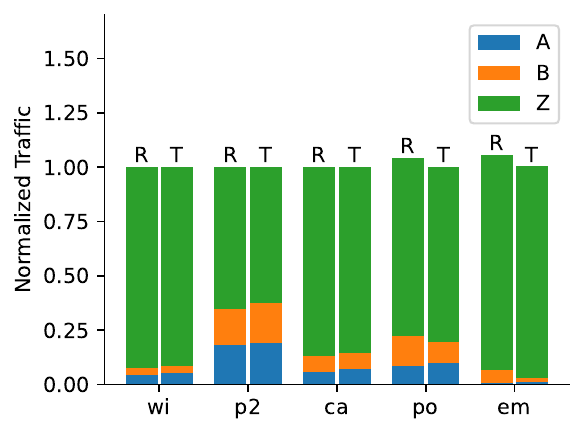}
  \vspace{-0.2in}
  \caption{Gamma Memory Traffic}
  \label{fig:eval:mem:gamma}
\end{subfigure}
\hfill
\begin{subfigure}{.3\textwidth}
  \centering 
  \includegraphics[width=\linewidth]{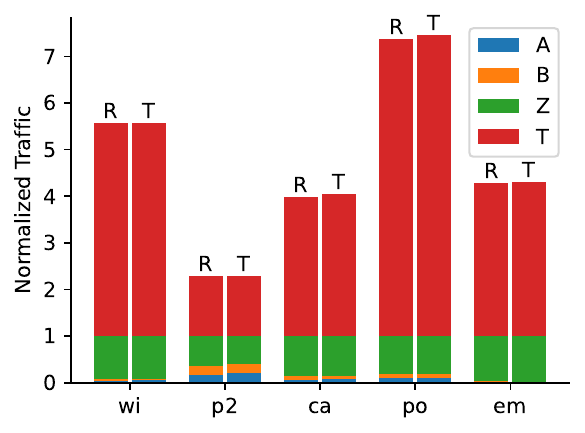}
  \vspace{-0.2in}
  \caption{OuterSPACE Memory Traffic}
  \label{fig:eval:mem:outerspace}
\end{subfigure}
\vspace{-0.05in}
\caption{\small Comparison of the memory traffic reported by the original publication (left, ``R'') and the \specname model (right ``T'') for ExTensor, Gamma, and OuterSPACE on 5 real-world tensors, discussed in Section~\ref{sec:eval:validation}. Traffic is normalized to the algorithmic minimum, and
benchmark acronyms are defined in Table~\ref{spec:tensors}. 
Tensor names $A$, $B$, $T$, and $Z$ correspond to the names defined in the Einsums in Figures~\ref{fig:outerspace:spec}~and~\ref{spec:accelerators}. 
$PO$ denotes the partial outputs (i.e. $Z$ traffic that is not the final write of $Z$).
}
\label{fig:eval:mem}
\end{figure*}

\begin{figure*}[ht]
\begin{subfigure}[c]{0.38\columnwidth}
  \centering
  \includegraphics[width=\columnwidth]{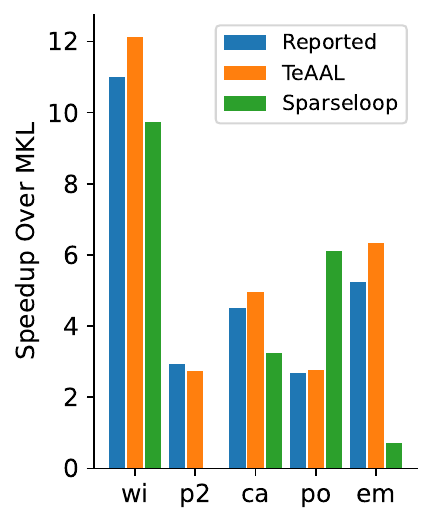}
  \vspace{-0.2in}
  \caption{ExTensor}
  \label{fig:eval:time:extensor}
\end{subfigure}
\hfill
\begin{subfigure}[c]{.38\columnwidth}
  \centering
  \includegraphics[width=\columnwidth]{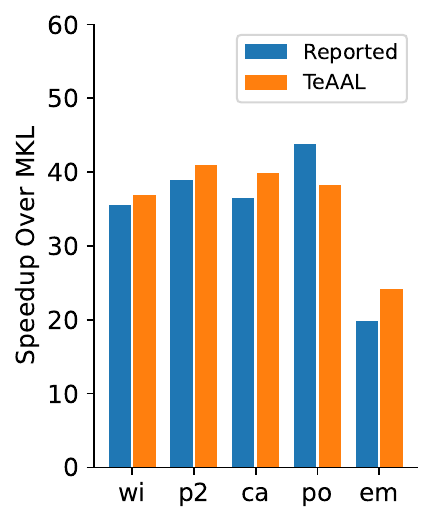}
  \vspace{-0.2in}
  \caption{Gamma}
  \label{fig:eval:time:gamma}
\end{subfigure}
\hfill
\begin{subfigure}[c]{0.58\columnwidth}
  \centering
  \includegraphics[width=\columnwidth]{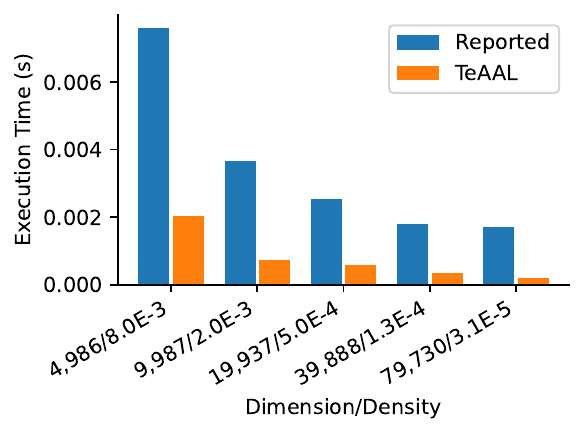}
  \vspace{-0.2in}
  \caption{OuterSPACE}
  \label{fig:eval:time:outerspace}
\end{subfigure}
\hfill
\begin{subfigure}[c]{.58\columnwidth}
  \centering
  \includegraphics[width=\columnwidth]{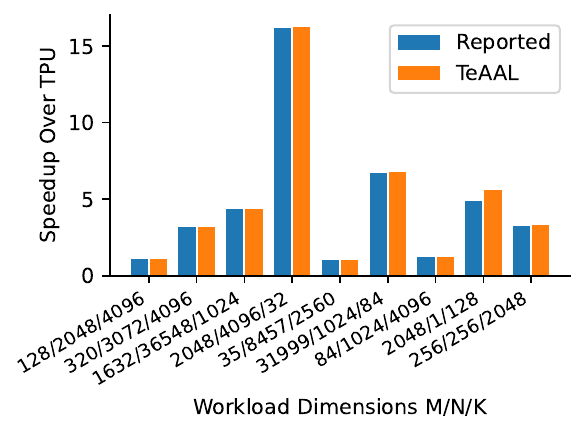}
  \vspace{-0.2in}
  \caption{SIGMA}
  \label{fig:eval:time:sigma}
\end{subfigure}
\vspace{-0.05in}
\caption{\small Validation of \specname-generated timing models against numbers/simulators reported/used in the original publications, also discussed in Section~\ref{sec:eval:validation}.
Benchmark acronyms are given in Table~\ref{spec:tensors}.
}
\label{fig:eval:time}
\vspace{-0.05in}
\end{figure*}

\begin{figure}[th]
\centering
\includegraphics[width=0.75\columnwidth]{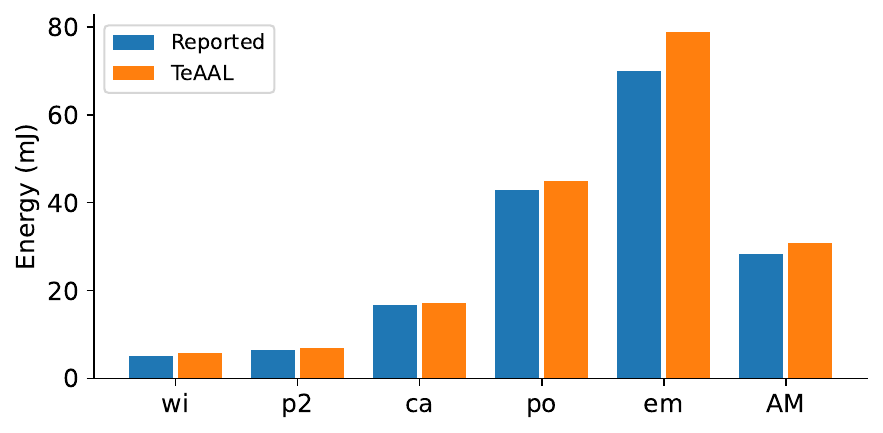}
\vspace{-0.1in}
\caption{Validation of the ExTensor energy model, also discussed in Section~\ref{sec:eval:validation}. Benchmark acronyms can be found in Table~\ref{spec:tensors}.
}
\label{fig:eval:energy}
\vspace{-0.15in}
\end{figure}

In this section, we describe a set of experiments used to validate \specname as an accurate cost model. 
Specifically, we compare memory traffic, performance, and power as reported by \specname to the numbers reported in the papers originally proposing each accelerator.
We report all averages as arithmetic means, following the methodology presented in \cite{perf_stats}.

\textbf{Memory Traffic.} Figure \ref{fig:eval:mem} presents a comparison of the memory traffic of the \specname models of each of the accelerators to the corresponding baseline. 
We use the first five tensors in Table~\ref{spec:tensors} because the prior work evaluates these tensors.
The takeaway is that we can reproduce each accelerator's memory traffic with low error (on average, $3.8\%$).
The single outlier, ExTensor on $\mathsf{p2}$, is caused by slightly different policies for eager loading between ExTensor and \specname (Section~\ref{sec:model:specs:binding}).
This policy difference is not fundamental, and can be remedied with additional effort.
We were unable to validate \specname's memory traffic model of SIGMA because there were no baseline numbers available.

\textbf{Performance.} 
Figure~\ref{fig:eval:time} presents a performance comparison of each \specname model against the reported numbers and 
Sparseloop~\cite{sparseloop_micro}'s estimate, when possible.
We evaluate on the same five tensors as were used in the memory-traffic study or uniformly random sparse tensors.

Figures~\ref{fig:eval:time:extensor} and \ref{fig:eval:time:gamma} show the speedup of ExTensor and Gamma, respectively, over Intel MKL.
\specname shows consistently low error rates for each (on average, $9.0\%$ and $6.6\%$, respectively).
We perform an analogous evaluation for SIGMA in  Figure~\ref{fig:eval:time:sigma}, relative to a Google Cloud TPU and using 
synthetic matrices with uniform-random sparsity (where all matrices $A$ and $B$ had 80\% and 10\% sparsity, respectively).
Here, we show an average error of only 2.5\%.

We compare Sparseloop's results to ours on ExTensor in Figure~\ref{fig:eval:time:extensor}.\footnote{We were unable to use Sparseloop to model Gamma/OuterSPACE (because it does not support cascades of Einsums) or SIGMA (because its occupancy-based partitioning is not sufficiently general).}
Sparseloop has an average error of $187\%$, which we attribute to its analytical sparsity distribution.
We could not model $\mathsf{p2}$ on Sparseloop because the tensor shape caused an integer overflow error.

Because we could not obtain the raw baseline numbers used in the OuterSPACE paper,
Figure~\ref{fig:eval:time:outerspace} shows the performance of the original simulator and the \specname model on a number of synthetic sparse matrices generated from a uniform-random distribution.
On average, our cost model is consistently $\sim 80\%$ faster than the original simulator, though the overall trend is consistent.
In Figure~\ref{fig:eval:mem:outerspace}, we showed that \specname models OuterSPACE's memory traffic with a $<1.8\%$ error, so we suspect that the discrepancy in execution time comes from an undocumented (and, therefore, unmodeled) feature of the OuterSPACE PE microarchitecture.

\textbf{Energy.} Figure~\ref{fig:eval:energy} compares \specname's energy estimates to the reported baseline, showing consistently low error rates---on average, 7.8\%.
\specname over estimates energy on $\mathsf{em}$ because it overestimates ExTensor's memory traffic on that benchmark (see Figure~\ref{fig:eval:mem:extensor}).
Since none of the other accelerators reported their per-component, per-action energy consumption characteristics, we were unable to validate the power model on those designs.
\section{Improving Graphicionado}
\label{sec:eval:design}

\begin{figure}
\vspace{-0.1in}
\begin{subfigure}{0.37\columnwidth}
\centering
\begin{lstlisting}[basicstyle=\small, commentstyle=\color{teal}]
# Processing Phase
- SO[d, s] = take(G[d, s], A0[s], 0) |\label{line:g:so}|
- R[d] = SO[d, s] * A0[s] |\label{line:g:r}|

# Apply Phase
- P1[v] = R[v] + P0[v] |\label{line:g:p1}|
- M[v] = P1[v] - P0[v] |\label{line:g:m}|
- A1[v] = take(M[v], P1[v], 1) |\label{line:g:a1}|
\end{lstlisting}
\vspace{-0.05in}
\caption{\small 
Graphicionado~\cite{graphicionado}
}
\label{fig:novel:graphicionado}
\end{subfigure} \hfill
\begin{subfigure}{0.38\columnwidth}
\centering
\begin{lstlisting}[basicstyle=\small, commentstyle=\color{teal}]
# Processing Phase
- SO[d, s] = take(G[d, s], A0[s], 0)
- R[d] = SO[d, s] * A0[s]

# Apply Phase
- MP[v] = take(R[v], P0[v], 1) |\label{line:sa:mp}|
- NP[v] = R[v] + MP[v]
- M[v] = NP[v] - MP[v]
- P0[v] = take(M[v], NP[v], 1) |\label{line:sa:p1}|
- A1[v] = take(M[v], NP[v], 1)
- P1 = P0
\end{lstlisting}
\vspace{-0.05in}
\caption{\small GraphDynS~\cite{graphdyns}}
\label{fig:novel:sparse_access}
\end{subfigure}\hfill
\vspace{-0.05in}
\caption{\small Einsum cascades for two vertex-centric programming accelerators. A specific algorithm manifests by redefining the $\times$ and $+$ operators (e.g., for SSSP, to addition and minimum, respectively).
}
\label{fig:novel:specs}
\end{figure}

\begin{figure*}[t]
\begin{subfigure}[T]{.3\textwidth}
  \centering
  \includegraphics[width=\linewidth]{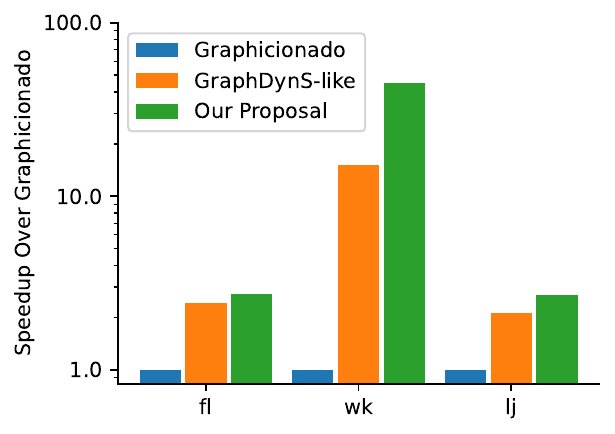}
  \vspace{-0.2in}
  \caption{\small Speed-Up on BFS}
  \label{fig:eval:novel:bfs}
\end{subfigure}
\hfill
\begin{subfigure}[T]{.3\textwidth}
  \centering
  \includegraphics[width=\linewidth]{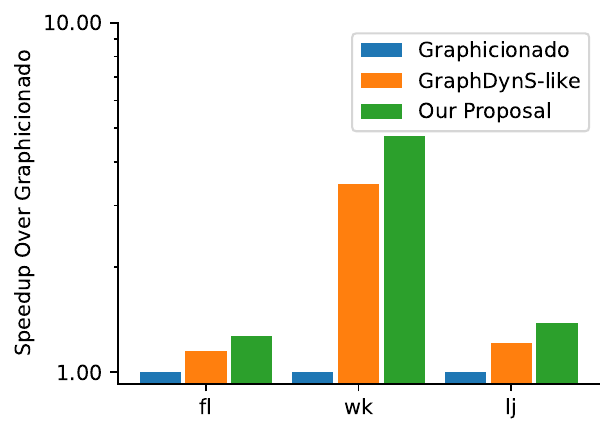}  \vspace{-0.2in}
  \caption{\small Speed-Up on SSSP}
  \label{fig:eval:novel:sssp}
\end{subfigure}
\hfill
\begin{subfigure}[T]{.3\textwidth}
  \centering
  \includegraphics[width=\linewidth]{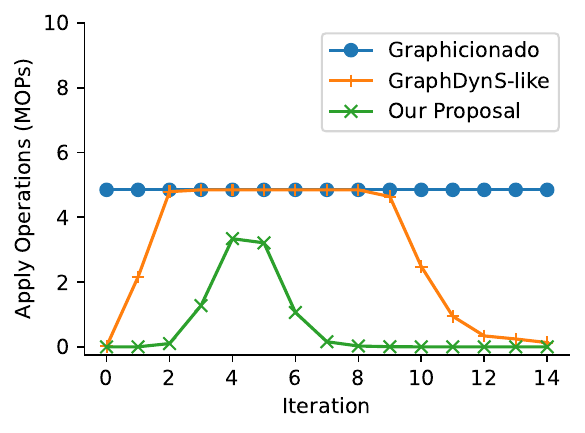}  \vspace{-0.2in}
  \caption{\small Apply Ops Performed for $\mathsf{lj}$ on BFS}
  \label{fig:eval:novel:ops}
\end{subfigure}
\vspace{-0.1in}
\caption{\small Comparison between Graphicionado and improved accelerators. All designs use Graphicionado's parameterization (Table~\ref{spec:accel-hw}).
}
\label{fig:eval:novel}
\vspace{-0.1in}
\end{figure*}

In this section, we demonstrate both the generality of \specname as a tool for modeling a broader class of accelerators and its value when proposing and evaluating new designs.
As an example, we model Grapicionado~\cite{graphicionado}---an accelerator for graph algorithms written in the vertex-centric programming paradigm---and GraphDynS~\cite{graphdyns}---an accelerator that optimizes this design.
Finally, we propose further optimizations to GraphDynS and demonstrate their value by evaluating all three designs using \specname.

Vertex-centric programming describes graph algorithms from the point of view of a single vertex.
During each iteration, a vertex, if active, sends its property to all of its destination vertices.
Then, the vertex, whether or not it is active, processes its incoming properties, reduces them to a single new value, and applies this value to its property~\cite{graphmat}.
In this evaluation, we will focus on algorithms where a subset of the vertices are active each iteration.

Figure~\ref{fig:novel:graphicionado} shows a cascade of Einsums representing Graphicionado~\cite{graphicionado}.
We omit the rest of the \specname specification for space.
Graphicionado divides its evaluation into two stages.
During the processing phase, the active vertices $A0$ are used to select the edges that need to be processed $SO$ (Line~\ref{line:g:so}), the weights of those edges are combined with the source vertex properties (Line~\ref{line:g:r}), and reduced into $R$ (implicit in Line~\ref{line:g:r}).
Then, during the apply phase, the vertex property $P0$/$P1$ is updated (Line~\ref{line:g:p1}) and the new set of active vertices $A1$ is created using a mask $M$ of updated vertices (Lines~\ref{line:g:m}-\ref{line:g:a1}).
By redefining the multiplication and addition operators (e.g., for single source shortest path (SSSP), to addition and minimum, respectively), this represents a functionally correct implementation of a graph kernel written in the vertex-centric programming model.
Through private correspondence with the Graphicionado authors, we found that all data, simulators, etc. used in this paper are proprietary, making it impossible for us to perform a similar analysis to Section~\ref{sec:eval:validation}.
However, using \specname, we were able to profile Graphicionado ourselves and compare it with other designs.

GraphDynS~\cite{graphdyns} optimizes Graphicionado by adding new Einsums to the cascade to take advantage of the sparsity of $R$.
Figure~\ref{fig:novel:sparse_access} shows the updated cascade.
Building an additional intermediate $MP$ (Line~\ref{line:sa:mp}), containing the values of $P0$ that can be modified, decreases the memory traffic incurred by $P0$ and the number of apply operations the accelerator needs to perform.
Filtering the writes to $P0$ with $M$ (Line~\ref{line:sa:p1}) also decreases the memory traffic.
GraphDynS implements this optimization by keeping a 256-element bitmap, where each bit corresponds to $1 / 256$th of the vertices.
In \specname, this manifests as an additional $\mathsf{uniform\_shape}$ partitioning.
If the bit is 1, the accelerator eagerly loads the entire partition of vertex properties.
GraphDynS further improves upon Graphicionado by changing the format of the graph from an edge-list representation to CSR.
This format change eliminates unnecessary reloading of the source vertex ID and removes the loading of the edge weight for algorithms that do not use it (e.g., BFS).

We optimize this design by removing the partitioning, allowing us to only load the property and perform the apply on vertices that are actually modified during the processing phase.
Our proposed accelerator also implements the format optimization.

We evaluate Graphicionado, GraphDynS, and our proposal on a subset of the graphs and all sparse active vertex set algorithms evaluated in the original Graphiciondo paper.
To enable an apples-to-apples comparison, we use the hardware parameters chosen for Graphicionado (see Table~\ref{spec:accel-hw}).
Figure~\ref{fig:eval:novel} shows the speedup achieved by each of the designs over Graphicionado.
Figure~\ref{fig:eval:novel:bfs} shows that our proposal enables an average of $1.9\times$ improvement over GraphDynS on BFS, while Figure~\ref{fig:eval:novel:sssp} shows that our proposal enables an average of $1.2\times$ improvement over GraphDynS on SSSP.
Figure~\ref{fig:eval:novel:ops} explains this improvement.
While GraphDynS's bitmap approach reduces the number of apply operations required when the set of active vertices is small, our proposal is also able to skip apply operations when the set of active vertices is large.

This study shows that \specname can express designs in domains beyond sparse tensor algebra.
Furthermore, it also demonstrates \specname's value as a tool for qualitatively and quantitatively comparing designs, improving the iterative design refinement process.
Notably, our proposed optimization only required meaningful changes to the mapping specification.
By decomposing the design of an accelerator into a hierarchy of specifications, \specname enables us to efficiently express existing designs and propose new optimizations.
\section{Related Work}

{\renewcommand{\arraystretch}{1.2}
\setlength{\tabcolsep}{5pt}

\newcommand{\scc}{\small\checkmark}

\begin{table}[t]
\centering
\footnotesize
\caption{\label{related:compare:sparse} \small 
Sparse tensor modeling frameworks.
}
\vspace{-0.05in}
\begin{tabular}{ M{2.2cm} | M{.85cm} | M{1.1cm} | M{0.4cm} | M{.7cm} | M{1.025cm} }
      & \textbf{STONNE} \newline \cite{stonne, flexagon} & \textbf{Sparseloop}\newline\cite{sparseloop_micro} &
     \textbf{SAM}\newline\cite{sam} &
     \textbf{CIN-P} \newline \cite{ahrens:2021:aac} & 
     \textbf{\specname} \newline (\emph{this work})  \\
    \hline \hline
    \textbf{Models Hardware}        & \scc   & \scc      & \scc &   & \scc \\ \hline
    \textbf{Generic Kernels}        &        & \scc  & \scc & \scc  & \scc \\ \hline
    \textbf{Cascaded Einsums}       &        &       & \scc & \scc  & \scc \\ \hline
    \textbf{Index Expressions}      &        & \scc  &      &  & \scc \\ \hline
    \textbf{Shape-Based Part.}      &        & \scc  & \scc &  & \scc \\ \hline
    \textbf{Occ.-Based Part.}       &        &       & \scc &  & \scc \\ \hline
    \textbf{Generic Flattening}     &        &       & \scc &  & \scc \\ \hline
    \textbf{Rank Swizzling}         &        &       &      & \scc & \scc \\ \hline
    \textbf{Format Expressivity}    &        & \scc  & \scc & \scc & \scc \\ \hline
    \textbf{Caches}                 & \scc   &       &      &  & \scc \\ \hline
    \textbf{Precise Data Set}       & \scc   &       & \scc &  & \scc \\ \hline
    \textbf{High Model Fidelity}    & \scc   &       &      &  & \scc \\
\end{tabular}
\vspace{-0.2in}
\end{table}

The rise in machine learning and tensor algebra accelerators has been followed by an increase in  
tools that explore the accelerator design space and model 
various efficiency characteristics~\cite{interstellar, timeloop, sparseloop_micro, maestro, senanayake:2020:asi, cosa, zigzag, stonne, ahrens:2021:aac}.
Most frameworks solely support dense computations and target DNN applications~\cite{timeloop, maestro, cosa, zigzag}.

Table~\ref{related:compare:sparse} compares frameworks that model architectures computing on sparse tensors.
STONNE~\cite{stonne} is a cycle-level modeling framework for DNN accelerators. 
Like \specname, STONNE's analysis is data-driven; however, the only sparse workload it supports is SpMSpM.

Two other works---Sparseloop~\cite{sparseloop_micro} and the Sparse Abstract Machine (SAM)~\cite{sam}---model sparse workloads expressed in the Einsum language, but with lower fidelity than \specname. 
Sparseloop~\cite{sparseloop_micro} has a flexible hardware backend and takes as input a specification of the architecture, a statistical model of the data, sparse optimizations such as intersection~\cite{extensor}, and a user-specified mapping. 
It returns estimates of performance and energy consumption. 
Unlike \specname, it does not support many of the features important to sparse computations, such as cascaded Einsums, caches, rank swizzling, and others.
Of note, Sparseloop models sparsity analytically using probability distributions.
By contrast, \specname evaluates on real tensors directly.
This enables \specname to achieve higher-fidelity modeling, at the cost of increased simulator runtime.
SAM~\cite{sam} targets an architecture consisting of hardware modules similar to those supported by \specname.
However, rather than generating high-fidelity models of specific accelerator designs, it models the accelerator dataflow on the SAM hardware.

Beyond accelerators, prior work on CIN-P~\cite{ahrens:2021:aac} considers sparse workload modeling for programmable devices such as CPUs.
Specifically, CIN-P is a mapper language that, when combined with an asymptotic cost model and autoscheduler, generates mappings that can then be compiled using TACO~\cite{taco}. 
Since it uses asymptotic analysis, the mapper only considers a small subset of the space of mapping decisions, including loop order and fusion.

Beyond frameworks for modeling/evaluating sparse tensor algebra kernels, there is also a closely related line of work targeted at compiling these kernels for existing programmable devices~\cite{taco, taco_formats, taco_workspaces, mosaic, unified_convolution_framework}.
These works provide a similar set of algorithms and mappings, with feature sets that overlap \specname.
For example, Mosaic~\cite{mosaic} supports modeling generic kernels, but inherits the limitations of the existing TACO~\cite{taco} language. 
It does not support several idioms that \specname does, e.g., allowing affine expressions as tensor indices (only adding constraints to the shapes of ranks) and only supports 1D intermediates~\cite{taco_workspaces} (as opposed to generic cascades).
However, unlike \specname, it allows users to mix existing library calls with user-defined kernels.
\section{Conclusion and Future Work}

This paper presented \specname: a language and simulator generator for describing and evaluating sparse tensor accelerators. 
The key contribution is to demonstrate
how state-of-the-art sparse accelerators
can be represented as cascades of mapped Einsums and content-preserving transformations on the Einsums' constituent tensors.
From this observation, we propose the \specname language which enables designers to utilize and combine these concepts (for both the modeling of current and the designing of new accelerators) and a generator from this language to executable simulators that emulate fibertree operations (lowered onto concrete data representations and hardware units).
Beyond enabling a more efficient accelerator design process, \specname also provides the architecture community with a common language for comparing and discussing designs.

Using cascades of Einsums as the problem specification enables \specname to model accelerators in or near the space of tensor algebra, including for deep learning, tensor decomposition, and graph algorithms.
We think that this abstraction can be extended to support a wider range of workloads and accelerators, e.g., by adding support for iterative algorithms or non-linear functions.
Following the discussion in Section~\ref{sec:model:express}, other missing features can be added by augmenting other (specific) levels in \specname's pyramidal abstraction hierarchy. 
Our thesis is that many unsupported features will manifest as changes to one level (and for that matter, a lower level) in the hierarchy---which suggests they will be relatively easy to incorporate.  
For example, design changes to various types of hardware blocks (e.g., mergers, arithmetic units, memories) are representable as point changes to the architecture specification language (as opposed to changing the cascade of Einsums abstraction). 
We leave a full investigation to future work.

Another important direction for future work is to incorporate \specname into a flow for performing design space exploration.
Such flows are typically hierarchical, starting with an exploration of the full design space using low fidelity models and only then performing a more in-depth, accurate analysis of the remaining promising designs~\cite{hierarchical_dse}.
We view \specname, as it is described in this work, as a middle level in this hierarchy.
The simulators generated by \specname are higher fidelity than those produced by an analytical model like Sparseloop~\cite{sparseloop_micro} and much more efficiently described by an architect than raw RTL.
Then, the main technical challenge to solve is how to use \specname to \emph{automatically} explore a space of designs.
Specifically, allowing accelerators to be described as a cascade of Einsums (instead of a single Einsum~\cite{timeloop, sparseloop_micro}) means that we would like to explore multiple cascades for the same problem, which requires a way to efficiently list the various promising cascades for that problem.  This task, as far as we know, is open.

Finally, while the current \specname backend generates performance/energy models, we think future work could support other backends (e.g., for generating hardware directly).

\paragraph{Acknowledgments}
We thank the anonymous reviewers for their valuable feedback.
This research was partially funded by NSF grants 
8191902 and 1942888, DARPA grant HR0011-18-3-0007,  
and by a UIUC SURGE Fellowship and Microsoft Research PhD Fellowship.
We would like to thank a number of people for their help in enabling us to better validate/extend ExTensor, Gamma, OuterSPACE, SIGMA, and Graphicionado: Tae Jun Ham, Kartik Hegde, Tushar Krishna, Francisco Muñoz-Martinez, Eric Qin, Daniel Sanchez, Hanrui Wang, Lisa Wu Wills, Guowei Zhang, and Zhekai Zhang.
We would also like to thank Yannan (Nellie) Wu for help with Sparseloop, and 
Timor Averbuch, Alex Dicheva, John D. Owens, Yasmin Sarita, and Xinrui (Alice) Wu for many helpful discussions.
Finally, we thank Willow Ahrens and Jaeyeon Won for feedback on early versions of the manuscript.

\bibliographystyle{ACM-Reference-Format}
\bibliography{refs, architecture, chris, deep_learning, tensors}


\begin{thebibliography}{56}


\ifx \showCODEN    \undefined \def \showCODEN     #1{\unskip}     \fi
\ifx \showDOI      \undefined \def \showDOI       #1{#1}\fi
\ifx \showISBNx    \undefined \def \showISBNx     #1{\unskip}     \fi
\ifx \showISBNxiii \undefined \def \showISBNxiii  #1{\unskip}     \fi
\ifx \showISSN     \undefined \def \showISSN      #1{\unskip}     \fi
\ifx \showLCCN     \undefined \def \showLCCN      #1{\unskip}     \fi
\ifx \shownote     \undefined \def \shownote      #1{#1}          \fi
\ifx \showarticletitle \undefined \def \showarticletitle #1{#1}   \fi
\ifx \showURL      \undefined \def \showURL       {\relax}        \fi
\providecommand\bibfield[2]{#2}
\providecommand\bibinfo[2]{#2}
\providecommand\natexlab[1]{#1}
\providecommand\showeprint[2][]{arXiv:#2}

\bibitem[hif(2023)]%
        {hifiber}
 \bibinfo{year}{2023}\natexlab{}.
\newblock \bibinfo{title}{Fibertree Project}.
\newblock
  \bibinfo{howpublished}{\url{https://github.com/Fibertree-Project/fibertree}}.
\newblock


\bibitem[Ahrens et~al\mbox{.}(2022)]%
        {ahrens:2021:aac}
\bibfield{author}{\bibinfo{person}{Peter Ahrens}, \bibinfo{person}{Fredrik
  Kjolstad}, {and} \bibinfo{person}{Saman~P. Amarasinghe}.}
  \bibinfo{year}{2022}\natexlab{}.
\newblock \showarticletitle{Autoscheduling for sparse tensor algebra with an
  asymptotic cost model}. In \bibinfo{booktitle}{\emph{{PLDI}'22}}.
\newblock


\bibitem[Aktulga et~al\mbox{.}(2014)]%
        {tallskinny1}
\bibfield{author}{\bibinfo{person}{Hasan~Metin Aktulga}, \bibinfo{person}{Aydin
  Bulu{\c{c}}}, \bibinfo{person}{Samuel Williams}, {and} \bibinfo{person}{Chao
  Yang}.} \bibinfo{year}{2014}\natexlab{}.
\newblock \showarticletitle{Optimizing sparse matrix-multiple vectors
  multiplication for nuclear configuration interaction calculations}. In
  \bibinfo{booktitle}{\emph{IPDPS'14}}.
\newblock


\bibitem[Albericio et~al\mbox{.}(2016)]%
        {albericio:2016:cin}
\bibfield{author}{\bibinfo{person}{Jorge Albericio}, \bibinfo{person}{Patrick
  Judd}, \bibinfo{person}{Tayler~H. Hetherington}, \bibinfo{person}{Tor~M.
  Aamodt}, \bibinfo{person}{Natalie D.~Enright Jerger}, {and}
  \bibinfo{person}{Andreas Moshovos}.} \bibinfo{year}{2016}\natexlab{}.
\newblock \showarticletitle{Cnvlutin: Ineffectual-Neuron-Free Deep Neural
  Network Computing}. In \bibinfo{booktitle}{\emph{{ISCA}'16}}.
\newblock


\bibitem[Azad et~al\mbox{.}(2015)]%
        {triangle_counting_tensor_algebra}
\bibfield{author}{\bibinfo{person}{Ariful Azad}, \bibinfo{person}{Aydın
  Buluc}, {and} \bibinfo{person}{John Gilbert}.}
  \bibinfo{year}{2015}\natexlab{}.
\newblock \showarticletitle{Parallel Triangle Counting and Enumeration Using
  Matrix Algebra}. In \bibinfo{booktitle}{\emph{IPDPSW'15}}.
\newblock


\bibitem[Bansal et~al\mbox{.}(2023)]%
        {mosaic}
\bibfield{author}{\bibinfo{person}{Manya Bansal}, \bibinfo{person}{Olivia Hsu},
  \bibinfo{person}{Kunle Olukotun}, {and} \bibinfo{person}{Fredrik Kjolstad}.}
  \bibinfo{year}{2023}\natexlab{}.
\newblock \showarticletitle{Mosaic: An Interoperable Compiler for Tensor
  Algebra}. In \bibinfo{booktitle}{\emph{PLDI'23}}.
\newblock


\bibitem[Chen et~al\mbox{.}(2018)]%
        {tvm}
\bibfield{author}{\bibinfo{person}{Tianqi Chen}, \bibinfo{person}{Thierry
  Moreau}, \bibinfo{person}{Ziheng Jiang}, \bibinfo{person}{Haichen Shen},
  \bibinfo{person}{Eddie~Q. Yan}, \bibinfo{person}{Leyuan Wang},
  \bibinfo{person}{Yuwei Hu}, \bibinfo{person}{Luis Ceze},
  \bibinfo{person}{Carlos Guestrin}, {and} \bibinfo{person}{Arvind
  Krishnamurthy}.} \bibinfo{year}{2018}\natexlab{}.
\newblock \showarticletitle{{TVM: E}nd-to-End Optimization Stack for Deep
  Learning}. In \bibinfo{booktitle}{\emph{OSDI'18}}.
\newblock


\bibitem[Chen et~al\mbox{.}(2016)]%
        {eyeriss}
\bibfield{author}{\bibinfo{person}{Yu-Hsin Chen}, \bibinfo{person}{Joel Emer},
  {and} \bibinfo{person}{Vivienne Sze}.} \bibinfo{year}{2016}\natexlab{}.
\newblock \showarticletitle{Eyeriss: A Spatial Architecture for
  Energy-efficient Dataflow for Convolutional Neural Networks}. In
  \bibinfo{booktitle}{\emph{ISCA'16}}.
\newblock


\bibitem[Chou et~al\mbox{.}(2018)]%
        {taco_formats}
\bibfield{author}{\bibinfo{person}{Stephen Chou}, \bibinfo{person}{Fredrik
  Kjolstad}, {and} \bibinfo{person}{Saman Amarasinghe}.}
  \bibinfo{year}{2018}\natexlab{}.
\newblock \showarticletitle{Format Abstraction for Sparse Tensor Algebra
  Compilers}. In \bibinfo{booktitle}{\emph{OOPSLA'18}}.
\newblock


\bibitem[Cooley and Tukey(1965)]%
        {cooley-tukey-fft}
\bibfield{author}{\bibinfo{person}{James~W. Cooley} {and}
  \bibinfo{person}{John~W. Tukey}.} \bibinfo{year}{1965}\natexlab{}.
\newblock \showarticletitle{An Algorithm for the Machine Calculation of Complex
  Fourier Series}.
\newblock \bibinfo{journal}{\emph{Math. Comp.}} (\bibinfo{year}{1965}).
\newblock
\showISSN{00255718, 10886842}


\bibitem[Davis and Hu(2011)]%
        {florida}
\bibfield{author}{\bibinfo{person}{Timothy~A Davis} {and}
  \bibinfo{person}{Yifan Hu}.} \bibinfo{year}{2011}\natexlab{}.
\newblock \showarticletitle{The University of Florida sparse matrix
  collection}.
\newblock  (\bibinfo{year}{2011}).
\newblock


\bibitem[Dongen(2000)]%
        {graph_clustering}
\bibfield{author}{\bibinfo{person}{Stijn Dongen}.}
  \bibinfo{year}{2000}\natexlab{}.
\newblock \showarticletitle{Graph Clustering by Flow Simulation}.
\newblock \bibinfo{journal}{\emph{PhD thesis, Center for Math and Computer
  Science (CWI)}} (\bibinfo{year}{2000}).
\newblock


\bibitem[{Einstein}(1916)]%
        {einsum}
\bibfield{author}{\bibinfo{person}{A. {Einstein}}.}
  \bibinfo{year}{1916}\natexlab{}.
\newblock \showarticletitle{The Foundation of the General Theory of
  Relativity}.
\newblock \bibinfo{journal}{\emph{Annalen der Physik}} (\bibinfo{year}{1916}).
\newblock


\bibitem[Ham et~al\mbox{.}(2016)]%
        {graphicionado}
\bibfield{author}{\bibinfo{person}{Tae~Jun Ham}, \bibinfo{person}{Lisa Wu},
  \bibinfo{person}{Narayanan Sundaram}, \bibinfo{person}{Nadathur Satish},
  {and} \bibinfo{person}{Margaret Martonosi}.} \bibinfo{year}{2016}\natexlab{}.
\newblock \showarticletitle{Graphicionado: A high-performance and
  energy-efficient accelerator for graph analytics}. In
  \bibinfo{booktitle}{\emph{MICRO'16}}.
\newblock


\bibitem[Han et~al\mbox{.}(2016)]%
        {eie}
\bibfield{author}{\bibinfo{person}{Song Han}, \bibinfo{person}{Xingyu Liu},
  \bibinfo{person}{Huizi Mao}, \bibinfo{person}{Jing Pu},
  \bibinfo{person}{Ardavan Pedram}, \bibinfo{person}{Mark~A Horowitz}, {and}
  \bibinfo{person}{William~J Dally}.} \bibinfo{year}{2016}\natexlab{}.
\newblock \showarticletitle{EIE: efficient inference engine on compressed deep
  neural network}. In \bibinfo{booktitle}{\emph{ISCA'16}}.
\newblock


\bibitem[Hegde et~al\mbox{.}(2019)]%
        {extensor}
\bibfield{author}{\bibinfo{person}{Kartik Hegde}, \bibinfo{person}{Hadi
  Asghari-Moghaddam}, \bibinfo{person}{Michael Pellauer}, \bibinfo{person}{Neal
  Crago}, \bibinfo{person}{Aamer Jaleel}, \bibinfo{person}{Edgar Solomonik},
  \bibinfo{person}{Joel Emer}, {and} \bibinfo{person}{Christopher~W.
  Fletcher}.} \bibinfo{year}{2019}\natexlab{}.
\newblock \showarticletitle{{ExTensor: A}n Accelerator for Sparse Tensor
  Algebra}. In \bibinfo{booktitle}{\emph{MICRO'19}}.
\newblock


\bibitem[Hegde et~al\mbox{.}(2021)]%
        {mindmappings}
\bibfield{author}{\bibinfo{person}{Kartik Hegde}, \bibinfo{person}{Po-An Tsai},
  \bibinfo{person}{Sitao Huang}, \bibinfo{person}{Vikas Chandra},
  \bibinfo{person}{Angshuman Parashar}, {and} \bibinfo{person}{Christopher~W.
  Fletcher}.} \bibinfo{year}{2021}\natexlab{}.
\newblock \showarticletitle{{Mind Mappings: E}nabling Efficient
  Algorithm-Accelerator Mapping Space Search}. In
  \bibinfo{booktitle}{\emph{ASPLOS'21}}.
\newblock


\bibitem[Hsu et~al\mbox{.}(2023)]%
        {sam}
\bibfield{author}{\bibinfo{person}{Olivia Hsu}, \bibinfo{person}{Maxwell
  Strange}, \bibinfo{person}{Ritvik Sharma}, \bibinfo{person}{Jaeyeon Won},
  \bibinfo{person}{Kunle Olukotun}, \bibinfo{person}{Joel~S. Emer},
  \bibinfo{person}{Mark~A. Horowitz}, {and} \bibinfo{person}{Fredrik
  Kj{\o}lstad}.} \bibinfo{year}{2023}\natexlab{}.
\newblock \showarticletitle{The Sparse Abstract Machine}. In
  \bibinfo{booktitle}{\emph{ASPLOS'23}}.
\newblock


\bibitem[Huang et~al\mbox{.}(2021)]%
        {cosa}
\bibfield{author}{\bibinfo{person}{Qijing Huang}, \bibinfo{person}{Minwoo
  Kang}, \bibinfo{person}{Grace Dinh}, \bibinfo{person}{Thomas Norell},
  \bibinfo{person}{Aravind Kalaiah}, \bibinfo{person}{James Demmel},
  \bibinfo{person}{John Wawrzynek}, {and} \bibinfo{person}{Yakun~Sophia Shao}.}
  \bibinfo{year}{2021}\natexlab{}.
\newblock \showarticletitle{{CoSA}: Scheduling by Constrained Optimization for
  Spatial Accelerators}. In \bibinfo{booktitle}{\emph{ISCA'21}}.
\newblock


\bibitem[Hutter et~al\mbox{.}(2014)]%
        {hutter:2014:cas}
\bibfield{author}{\bibinfo{person}{Jürg Hutter}, \bibinfo{person}{Marcella
  Iannuzzi}, \bibinfo{person}{Florian Schiffmann}, {and} \bibinfo{person}{Joost
  VandeVondele}.} \bibinfo{year}{2014}\natexlab{}.
\newblock \showarticletitle{{CP2K}: Atomistic simulations of condensed matter
  systems}.
\newblock \bibinfo{journal}{\emph{WIREs Computational Molecular Science}}
  (\bibinfo{year}{2014}).
\newblock


\bibitem[Jacob and Mudge(1995)]%
        {perf_stats}
\bibfield{author}{\bibinfo{person}{Bruce Jacob} {and}
  \bibinfo{person}{Trevor~N. Mudge}.} \bibinfo{year}{1995}\natexlab{}.
\newblock \showarticletitle{Notes on Calculating Computer Performance}.
\newblock


\bibitem[Kjolstad et~al\mbox{.}(2019)]%
        {taco_workspaces}
\bibfield{author}{\bibinfo{person}{Fredrik Kjolstad}, \bibinfo{person}{Peter
  Ahrens}, \bibinfo{person}{Shoaib Kamil}, {and} \bibinfo{person}{Saman
  Amarasinghe}.} \bibinfo{year}{2019}\natexlab{}.
\newblock \showarticletitle{Tensor Algebra Compilation with Workspaces}. In
  \bibinfo{booktitle}{\emph{CGO'19}}.
\newblock


\bibitem[Kjolstad et~al\mbox{.}(2017)]%
        {taco}
\bibfield{author}{\bibinfo{person}{Fredrik Kjolstad}, \bibinfo{person}{Shoaib
  Kamil}, \bibinfo{person}{Stephen Chou}, \bibinfo{person}{David Lugato}, {and}
  \bibinfo{person}{Saman Amarasinghe}.} \bibinfo{year}{2017}\natexlab{}.
\newblock \showarticletitle{The Tensor Algebra Compiler}. In
  \bibinfo{booktitle}{\emph{OOPSLA'17}}.
\newblock


\bibitem[Kumar and Davidson(1980)]%
        {hierarchical_dse}
\bibfield{author}{\bibinfo{person}{B. Kumar} {and} \bibinfo{person}{E.~S.
  Davidson}.} \bibinfo{year}{1980}\natexlab{}.
\newblock \showarticletitle{Computer System Design Using a Hierarchical
  Approach to Performance Evaluation}.
\newblock \bibinfo{journal}{\emph{CACM'80}} (\bibinfo{year}{1980}).
\newblock


\bibitem[Kwon et~al\mbox{.}(2019)]%
        {maestro}
\bibfield{author}{\bibinfo{person}{Hyoukjun Kwon}, \bibinfo{person}{Michael
  Pellauer}, {and} \bibinfo{person}{Tushar Krishna}.}
  \bibinfo{year}{2019}\natexlab{}.
\newblock \showarticletitle{{Understanding Reuse, Performance, and Hardware
  Cost of DNN Dataflows}: A Data-Centric Approach Using MAESTRO}. In
  \bibinfo{booktitle}{\emph{MICRO'19}}.
\newblock


\bibitem[Leskovec and Krevl(2014)]%
        {snapnets}
\bibfield{author}{\bibinfo{person}{Jure Leskovec} {and} \bibinfo{person}{Andrej
  Krevl}.} \bibinfo{year}{2014}\natexlab{}.
\newblock \bibinfo{title}{{SNAP Datasets}: {Stanford} Large Network Dataset
  Collection}.
\newblock \bibinfo{howpublished}{\url{http://snap.stanford.edu/data}}.
\newblock


\bibitem[Mahmoud et~al\mbox{.}(2020)]%
        {mahmoud:2020:tde}
\bibfield{author}{\bibinfo{person}{Mostafa Mahmoud}, \bibinfo{person}{Isak
  Edo}, \bibinfo{person}{Ali~Hadi Zadeh}, \bibinfo{person}{Omar~Mohamed Awad},
  \bibinfo{person}{Gennady Pekhimenko}, \bibinfo{person}{Jorge Albericio},
  {and} \bibinfo{person}{Andreas Moshovos}.} \bibinfo{year}{2020}\natexlab{}.
\newblock \showarticletitle{{TensorDash}: Exploiting Sparsity to Accelerate
  Deep Neural Network Training}. In \bibinfo{booktitle}{\emph{MICRO'20}}.
\newblock


\bibitem[Mattson et~al\mbox{.}(2013)]%
        {graphblas}
\bibfield{author}{\bibinfo{person}{Tim Mattson}, \bibinfo{person}{David Bader},
  \bibinfo{person}{Jon Berry}, \bibinfo{person}{Aydin Buluc},
  \bibinfo{person}{Jack Dongarra}, \bibinfo{person}{Christos Faloutsos},
  \bibinfo{person}{John Feo}, \bibinfo{person}{John Gilbert},
  \bibinfo{person}{Joseph Gonzalez}, \bibinfo{person}{Bruce Hendrickson},
  \bibinfo{person}{Jeremy Kepner}, \bibinfo{person}{Charles Leiserson},
  \bibinfo{person}{Andrew Lumsdaine}, \bibinfo{person}{David Padua},
  \bibinfo{person}{Stephen Poole}, \bibinfo{person}{Steve Reinhardt},
  \bibinfo{person}{Mike Stonebraker}, \bibinfo{person}{Steve Wallach}, {and}
  \bibinfo{person}{Andrew Yoo}.} \bibinfo{year}{2013}\natexlab{}.
\newblock \showarticletitle{Standards for graph algorithm primitives}. In
  \bibinfo{booktitle}{\emph{HPEC'13}}.
\newblock


\bibitem[Mei et~al\mbox{.}(2020)]%
        {zigzag}
\bibfield{author}{\bibinfo{person}{Linyan Mei}, \bibinfo{person}{Pouya
  Houshmand}, \bibinfo{person}{Vikram Jain}, \bibinfo{person}{Sebastian
  Giraldo}, {and} \bibinfo{person}{Marian Verhelst}.}
  \bibinfo{year}{2020}\natexlab{}.
\newblock \showarticletitle{{ZigZag}: A Memory-Centric Rapid {DNN} Accelerator
  Design Space Exploration Framework}. In \bibinfo{booktitle}{\emph{Arxiv'20}}.
\newblock


\bibitem[Mu{\~{n}}oz{-}Mart{\'{\i}}nez et~al\mbox{.}(2023)]%
        {flexagon}
\bibfield{author}{\bibinfo{person}{Francisco Mu{\~{n}}oz{-}Mart{\'{\i}}nez},
  \bibinfo{person}{Raveesh Garg}, \bibinfo{person}{Michael Pellauer},
  \bibinfo{person}{Jos{\'{e}}~L. Abell{\'{a}}n}, \bibinfo{person}{Manuel~E.
  Acacio}, {and} \bibinfo{person}{Tushar Krishna}.}
  \bibinfo{year}{2023}\natexlab{}.
\newblock \showarticletitle{Flexagon: {A} Multi-dataflow Sparse-Sparse Matrix
  Multiplication Accelerator for Efficient {DNN} Processing}. In
  \bibinfo{booktitle}{\emph{ASPLOS'23}}.
\newblock


\bibitem[Muñoz-Martínez et~al\mbox{.}(2021)]%
        {stonne}
\bibfield{author}{\bibinfo{person}{Francisco Muñoz-Martínez},
  \bibinfo{person}{José~L. Abellán}, \bibinfo{person}{Manuel~E. Acacio},
  {and} \bibinfo{person}{Tushar Krishna}.} \bibinfo{year}{2021}\natexlab{}.
\newblock \showarticletitle{STONNE: Enabling Cycle-Level Microarchitectural
  Simulation for DNN Inference Accelerators}. In
  \bibinfo{booktitle}{\emph{IISWC'21}}.
\newblock


\bibitem[Nagasaka et~al\mbox{.}(2019)]%
        {nagasaka2019performance}
\bibfield{author}{\bibinfo{person}{Yusuke Nagasaka}, \bibinfo{person}{Satoshi
  Matsuoka}, \bibinfo{person}{Ariful Azad}, {and} \bibinfo{person}{Ayd{\i}n
  Bulu{\c{c}}}.} \bibinfo{year}{2019}\natexlab{}.
\newblock \showarticletitle{Performance optimization, modeling and analysis of
  sparse matrix-matrix products on multi-core and many-core processors}.
\newblock \bibinfo{journal}{\emph{Parallel Comput.}} (\bibinfo{year}{2019}).
\newblock


\bibitem[Odemuyiwa et~al\mbox{.}(2023)]%
        {tactile}
\bibfield{author}{\bibinfo{person}{Toluwanimi~O. Odemuyiwa},
  \bibinfo{person}{Hadi Asghari-Moghaddam}, \bibinfo{person}{Michael Pellauer},
  \bibinfo{person}{Kartik Hegde}, \bibinfo{person}{Po-An Tsai},
  \bibinfo{person}{Neal Crago}, \bibinfo{person}{Aamer Jaleel},
  \bibinfo{person}{John~D. Owens}, \bibinfo{person}{Edgar Solomonik},
  \bibinfo{person}{Joel Emer}, {and} \bibinfo{person}{Christopher Fletcher}.}
  \bibinfo{year}{2023}\natexlab{}.
\newblock \showarticletitle{Accelerating Sparse Data Orchestration via Dynamic
  Reflexive Tiling}. In \bibinfo{booktitle}{\emph{ASPLOS'23}}.
\newblock


\bibitem[Pal et~al\mbox{.}(2018)]%
        {outerspace}
\bibfield{author}{\bibinfo{person}{Subhankar Pal}, \bibinfo{person}{Jonathan
  Beaumont}, \bibinfo{person}{Dong-Hyeon Park}, \bibinfo{person}{Aporva
  Amarnath}, \bibinfo{person}{Siying Feng}, \bibinfo{person}{Chaitali
  Chakrabarti}, \bibinfo{person}{Hun-Seok Kim}, \bibinfo{person}{David Blaauw},
  \bibinfo{person}{Trevor Mudge}, {and} \bibinfo{person}{Ronald Dreslinski}.}
  \bibinfo{year}{2018}\natexlab{}.
\newblock \showarticletitle{{OuterSPACE}: An Outer Product Based Sparse Matrix
  Multiplication Accelerator}. In \bibinfo{booktitle}{\emph{HPCA'18}}.
\newblock


\bibitem[Parashar et~al\mbox{.}(2019)]%
        {timeloop}
\bibfield{author}{\bibinfo{person}{Angshuman Parashar},
  \bibinfo{person}{Priyanka Raina}, \bibinfo{person}{Yakun~Sophia Shao},
  \bibinfo{person}{Yu-Hsin Chen}, \bibinfo{person}{Victor~A. Ying},
  \bibinfo{person}{Anurag Mukkara}, \bibinfo{person}{Rangharajan Venkatesan},
  \bibinfo{person}{Brucek Khailany}, \bibinfo{person}{Stephen~W. Keckler},
  {and} \bibinfo{person}{Joel Emer}.} \bibinfo{year}{2019}\natexlab{}.
\newblock \showarticletitle{Timeloop: A Systematic Approach to DNN Accelerator
  Evaluation}. In \bibinfo{booktitle}{\emph{ISPASS'19}}.
\newblock


\bibitem[Parashar et~al\mbox{.}(2017)]%
        {SCNN}
\bibfield{author}{\bibinfo{person}{Angshuman Parashar}, \bibinfo{person}{Minsoo
  Rhu}, \bibinfo{person}{Anurag Mukkara}, \bibinfo{person}{Antonio Puglielli},
  \bibinfo{person}{Rangharajan Venkatesan}, \bibinfo{person}{Brucek Khailany},
  \bibinfo{person}{Joel Emer}, \bibinfo{person}{Stephen~W Keckler}, {and}
  \bibinfo{person}{William~J Dally}.} \bibinfo{year}{2017}\natexlab{}.
\newblock \showarticletitle{{SCNN}: An accelerator for compressed-sparse
  convolutional neural networks}. In \bibinfo{booktitle}{\emph{ISCA'17}}.
\newblock


\bibitem[Pellauer et~al\mbox{.}(2019)]%
        {buffets}
\bibfield{author}{\bibinfo{person}{Michael Pellauer},
  \bibinfo{person}{Yakun~Sophia Shao}, \bibinfo{person}{Jason Clemons},
  \bibinfo{person}{Neal Crago}, \bibinfo{person}{Kartik Hegde},
  \bibinfo{person}{Rangharajan Venkatesan}, \bibinfo{person}{Stephen Keckler},
  \bibinfo{person}{Christopher~W. Fletcher}, {and} \bibinfo{person}{Joel
  Emer}.} \bibinfo{year}{2019}\natexlab{}.
\newblock \showarticletitle{{Buffets}: An Efficient and Composable Storage
  Idiom for Explicit Decoupled Data Orchestration}. In
  \bibinfo{booktitle}{\emph{ASPLOS'19}}.
\newblock


\bibitem[Qin et~al\mbox{.}(2020)]%
        {sigma}
\bibfield{author}{\bibinfo{person}{Eric Qin}, \bibinfo{person}{Ananda
  Samajdar}, \bibinfo{person}{Hyoukjun Kwon}, \bibinfo{person}{Vineet Nadella},
  \bibinfo{person}{Sudarshan Srinivasan}, \bibinfo{person}{Dipankar Das},
  \bibinfo{person}{Bharat Kaul}, {and} \bibinfo{person}{Tushar Krishna}.}
  \bibinfo{year}{2020}\natexlab{}.
\newblock \showarticletitle{{SIGMA}: A Sparse and Irregular GEMM Accelerator
  with Flexible Interconnects for DNN Training}. In
  \bibinfo{booktitle}{\emph{HPCA'20}}.
\newblock


\bibitem[Ragan-Kelley et~al\mbox{.}(2013)]%
        {halide}
\bibfield{author}{\bibinfo{person}{Jonathan Ragan-Kelley},
  \bibinfo{person}{Connelly Barnes}, \bibinfo{person}{Andrew Adams},
  \bibinfo{person}{Sylvain Paris}, \bibinfo{person}{Fr\'{e}do Durand}, {and}
  \bibinfo{person}{Saman Amarasinghe}.} \bibinfo{year}{2013}\natexlab{}.
\newblock \showarticletitle{{Halide}: A Language and Compiler for Optimizing
  Parallelism, Locality, and Recomputation in Image Processing Pipelines}. In
  \bibinfo{booktitle}{\emph{PLDI'13}}.
\newblock


\bibitem[Senanayake et~al\mbox{.}(2020)]%
        {senanayake:2020:asi}
\bibfield{author}{\bibinfo{person}{Ryan Senanayake}, \bibinfo{person}{Changwan
  Hong}, \bibinfo{person}{Ziheng Wang}, \bibinfo{person}{Amalee Wilson},
  \bibinfo{person}{Stephen Chou}, \bibinfo{person}{Shoaib Kamil},
  \bibinfo{person}{Saman Amarasinghe}, {and} \bibinfo{person}{Fredrik
  Kjolstad}.} \bibinfo{year}{2020}\natexlab{}.
\newblock \showarticletitle{A Sparse Iteration Space Transformation Framework
  for Sparse Tensor Algebra}. In \bibinfo{booktitle}{\emph{OOPSLA'20}}.
\newblock


\bibitem[Solomonik et~al\mbox{.}(2017)]%
        {tallskinny0}
\bibfield{author}{\bibinfo{person}{Edgar Solomonik}, \bibinfo{person}{Maciej
  Besta}, \bibinfo{person}{Flavio Vella}, {and} \bibinfo{person}{Torsten
  Hoefler}.} \bibinfo{year}{2017}\natexlab{}.
\newblock \showarticletitle{Scaling betweenness centrality using
  communication-efficient sparse matrix multiplication}. In
  \bibinfo{booktitle}{\emph{SC'17}}.
\newblock


\bibitem[Srivastava et~al\mbox{.}(2020a)]%
        {matraptor}
\bibfield{author}{\bibinfo{person}{Nitish Srivastava}, \bibinfo{person}{Hanchen
  Jin}, \bibinfo{person}{Jie Liu}, \bibinfo{person}{David Albonesi}, {and}
  \bibinfo{person}{Zhiru Zhang}.} \bibinfo{year}{2020}\natexlab{a}.
\newblock \showarticletitle{{MatRaptor}: A Sparse-Sparse Matrix Multiplication
  Accelerator Based on Row-Wise Product}. In
  \bibinfo{booktitle}{\emph{MICRO'20}}.
\newblock


\bibitem[Srivastava et~al\mbox{.}(2020b)]%
        {tensaurus}
\bibfield{author}{\bibinfo{person}{Nitish Srivastava}, \bibinfo{person}{Hanchen
  Jin}, \bibinfo{person}{Shaden Smith}, \bibinfo{person}{Hongbo Rong},
  \bibinfo{person}{David Albonesi}, {and} \bibinfo{person}{Zhiru Zhang}.}
  \bibinfo{year}{2020}\natexlab{b}.
\newblock \showarticletitle{{Tensaurus}: A Versatile Accelerator for Mixed
  Sparse-Dense Tensor Computations}. In \bibinfo{booktitle}{\emph{HPCA'20}}.
\newblock


\bibitem[Sundaram et~al\mbox{.}(2015)]%
        {graphmat}
\bibfield{author}{\bibinfo{person}{Narayanan Sundaram},
  \bibinfo{person}{Nadathur Satish}, \bibinfo{person}{Md~Mostofa~Ali Patwary},
  \bibinfo{person}{Subramanya~R. Dulloor}, \bibinfo{person}{Michael~J.
  Anderson}, \bibinfo{person}{Satya~Gautam Vadlamudi},
  \bibinfo{person}{Dipankar Das}, {and} \bibinfo{person}{Pradeep Dubey}.}
  \bibinfo{year}{2015}\natexlab{}.
\newblock \showarticletitle{{GraphMat}: High Performance Graph Analytics Made
  Productive}. In \bibinfo{booktitle}{\emph{VLDB'15}}.
\newblock


\bibitem[Sze et~al\mbox{.}(2020)]%
        {sze:2020:epo}
\bibfield{author}{\bibinfo{person}{Vivienne Sze}, \bibinfo{person}{Yu{-}Hsin
  Chen}, \bibinfo{person}{Tien{-}Ju Yang}, {and} \bibinfo{person}{Joel~S.
  Emer}.} \bibinfo{year}{2020}\natexlab{}.
\newblock \bibinfo{booktitle}{\emph{Efficient Processing of Deep Neural
  Networks}}.
\newblock \bibinfo{publisher}{Springer}.
\newblock


\bibitem[VandeVondele et~al\mbox{.}(2012)]%
        {vande:2012:lss}
\bibfield{author}{\bibinfo{person}{Joost VandeVondele}, \bibinfo{person}{Urban
  Borštnik}, {and} \bibinfo{person}{Jürg Hutter}.}
  \bibinfo{year}{2012}\natexlab{}.
\newblock \showarticletitle{Linear Scaling Self-Consistent Field Calculations
  with Millions of Atoms in the Condensed Phase}.
\newblock \bibinfo{journal}{\emph{Journal of Chemical Theory and Computation}}
  (\bibinfo{year}{2012}).
\newblock


\bibitem[Wang et~al\mbox{.}(2021)]%
        {dstc}
\bibfield{author}{\bibinfo{person}{Yang Wang}, \bibinfo{person}{Chen Zhang},
  \bibinfo{person}{Zhiqiang Xie}, \bibinfo{person}{Cong Guo},
  \bibinfo{person}{Yunxin Liu}, {and} \bibinfo{person}{Jingwen Leng}.}
  \bibinfo{year}{2021}\natexlab{}.
\newblock \showarticletitle{Dual-side Sparse Tensor Core}. In
  \bibinfo{booktitle}{\emph{ISCA'21}}.
\newblock


\bibitem[Wijeratne et~al\mbox{.}(2021)]%
        {mttkrp-fpga}
\bibfield{author}{\bibinfo{person}{Sasindu Wijeratne},
  \bibinfo{person}{Rajgopal Kannan}, {and} \bibinfo{person}{Viktor Prasanna}.}
  \bibinfo{year}{2021}\natexlab{}.
\newblock \showarticletitle{Reconfigurable Low-latency Memory System for Sparse
  Matricized Tensor Times Khatri-Rao Product on FPGA}. In
  \bibinfo{booktitle}{\emph{HPEC'21}}.
\newblock


\bibitem[Wilhelm et~al\mbox{.}(2016)]%
        {wilhelm:2016:lsc}
\bibfield{author}{\bibinfo{person}{Jan Wilhelm}, \bibinfo{person}{Patrick
  Seewald}, \bibinfo{person}{Mauro Del~Ben}, {and} \bibinfo{person}{Jürg
  Hutter}.} \bibinfo{year}{2016}\natexlab{}.
\newblock \showarticletitle{Large-Scale Cubic-Scaling Random Phase
  Approximation Correlation Energy Calculations Using a Gaussian Basis}.
\newblock \bibinfo{journal}{\emph{Journal of Chemical Theory and Computation}}
  (\bibinfo{year}{2016}).
\newblock


\bibitem[Won et~al\mbox{.}(2023)]%
        {unified_convolution_framework}
\bibfield{author}{\bibinfo{person}{Jaeyeon Won}, \bibinfo{person}{Changwan
  Hong}, \bibinfo{person}{Charith Mendis}, \bibinfo{person}{Joel Emer}, {and}
  \bibinfo{person}{Saman Amarasinghe}.} \bibinfo{year}{2023}\natexlab{}.
\newblock \showarticletitle{Unified Convolution Framework: A compiler-based
  approach to support sparse convolutions}. In
  \bibinfo{booktitle}{\emph{MLSys'23}}.
\newblock


\bibitem[Wu et~al\mbox{.}(2019)]%
        {accelergy}
\bibfield{author}{\bibinfo{person}{Yannan~Nellie Wu}, \bibinfo{person}{Joel~S.
  Emer}, {and} \bibinfo{person}{Vivienne Sze}.}
  \bibinfo{year}{2019}\natexlab{}.
\newblock \showarticletitle{Accelergy: An Architecture-Level Energy Estimation
  Methodology for Accelerator Designs}. In
  \bibinfo{booktitle}{\emph{ICCAD'19}}.
\newblock


\bibitem[Wu et~al\mbox{.}(2022)]%
        {sparseloop_micro}
\bibfield{author}{\bibinfo{person}{Yannan~Nellie Wu}, \bibinfo{person}{Po-An
  Tsai}, \bibinfo{person}{Angshuman Parashar}, \bibinfo{person}{Vivienne Sze},
  {and} \bibinfo{person}{Joel~S. Emer}.} \bibinfo{year}{2022}\natexlab{}.
\newblock \showarticletitle{Sparseloop: An Analytical Approach To Sparse Tensor
  Accelerator Modeling}. In \bibinfo{booktitle}{\emph{MICRO'22}}.
\newblock


\bibitem[Yan et~al\mbox{.}(2019)]%
        {graphdyns}
\bibfield{author}{\bibinfo{person}{Mingyu Yan}, \bibinfo{person}{Xing Hu},
  \bibinfo{person}{Shuangchen Li}, \bibinfo{person}{Abanti Basak},
  \bibinfo{person}{Han Li}, \bibinfo{person}{Xin Ma}, \bibinfo{person}{Itir
  Akgun}, \bibinfo{person}{Yujing Feng}, \bibinfo{person}{Peng Gu},
  \bibinfo{person}{Lei Deng}, \bibinfo{person}{Xiaochun Ye},
  \bibinfo{person}{Zhimin Zhang}, \bibinfo{person}{Dongrui Fan}, {and}
  \bibinfo{person}{Yuan Xie}.} \bibinfo{year}{2019}\natexlab{}.
\newblock \showarticletitle{{Alleviating irregularity in graph analytics
  acceleration}: A hardware/software co-design approach}. In
  \bibinfo{booktitle}{\emph{MICRO'19}}.
\newblock


\bibitem[Yang et~al\mbox{.}(2020)]%
        {interstellar}
\bibfield{author}{\bibinfo{person}{Xuan Yang}, \bibinfo{person}{Mingyu Gao},
  \bibinfo{person}{Qiaoyi Liu}, \bibinfo{person}{Jeff Setter},
  \bibinfo{person}{Jing Pu}, \bibinfo{person}{Ankita Nayak},
  \bibinfo{person}{Steven Bell}, \bibinfo{person}{Kaidi Cao},
  \bibinfo{person}{Heonjae Ha}, \bibinfo{person}{Priyanka Raina},
  \bibinfo{person}{Christos Kozyrakis}, {and} \bibinfo{person}{Mark Horowitz}.}
  \bibinfo{year}{2020}\natexlab{}.
\newblock \showarticletitle{{Interstellar}: Using Halide's Scheduling Language
  to Analyze DNN Accelerators}. In \bibinfo{booktitle}{\emph{ASPLOS'20}}.
\newblock


\bibitem[Zhang et~al\mbox{.}(2021)]%
        {gamma}
\bibfield{author}{\bibinfo{person}{Guowei Zhang}, \bibinfo{person}{Nithya
  Attaluri}, \bibinfo{person}{Joel~S. Emer}, {and} \bibinfo{person}{Daniel
  Sanchez}.} \bibinfo{year}{2021}\natexlab{}.
\newblock \showarticletitle{Gamma: Leveraging Gustavson’s Algorithm to
  Accelerate Sparse Matrix Multiplication}. In
  \bibinfo{booktitle}{\emph{ASPLOS'21}}.
\newblock


\bibitem[Zhang et~al\mbox{.}(2020)]%
        {sparch}
\bibfield{author}{\bibinfo{person}{Zhekai Zhang}, \bibinfo{person}{Hanrui
  Wang}, \bibinfo{person}{Song Han}, {and} \bibinfo{person}{William~J. Dally}.}
  \bibinfo{year}{2020}\natexlab{}.
\newblock \showarticletitle{SpArch: Efficient Architecture for Sparse Matrix
  Multiplication}. In \bibinfo{booktitle}{\emph{HPCA'20}}.
\newblock


\end{thebibliography}

 \pagebreak

\appendix
\section{Artifact Appendix}
\label{sec:app_ae}

\subsection{Abstract}

In this artifact, we provide the source code for TeAAL, a simulator generator for sparse tensor algebra accelerators, as well as the scripts required to display the results of the simulation. For ease-of-use, we provide a Docker container and a set of Jupyter notebooks through which to run the experiments. This artifact can be evaluated on an x86-84 machine with 256 GB of memory and 75 GB of disk space.

\subsection{Artifact check-list (meta-information)}

{\small
\begin{itemize}
  \item {\bf Algorithm: }Automatic generation of sparse tensor algebra accelerator simulators
  \item {\bf Program: }Python, Sparseloop
  \item {\bf Run-time environment: } Docker, Jupyter
  \item {\bf Hardware: }An x86-64 machine with 256 GB of memory and 125 GB of disk space
  \item {\bf Output: }Plots generated from scripts
  \item {\bf Experiments: }Automatic generation of sparse tensor algebra accelerator simulators and execution of those simulators on specific benchmark data
  \item {\bf How much disk space required (approximately)?: }125 GB
  \item {\bf How much time is needed to prepare workflow (approximately)?: }$< 30$ minutes
  \item {\bf How much time is needed to complete experiments (approximately)?: }70 hours
  \item {\bf Publicly available?: } Yes
  \item {\bf Code licenses (if publicly available)?: }MIT
\end{itemize}
}

\subsection{Description - How to Access}

\textbf{Manually:} The artifact is hosted on Github at \url{https://github.com/FPSG-UIUC/micro23-teaal-artifact}.
Following the instructions in this repository will allow you to run specific experiments and nicely display the graphs.

\smallskip
\noindent \textbf{Via the MLCommons CM Interface:} It is also accessible through the MLCommons CM interface at \url{https://github.com/ctuning/cm-reproduce-research-projects/tree/main/script/reproduce-ieee-acm-micro2023-paper-8}.
This method provides less control over what experiments are executed.

\subsection{Installation}

\textbf{Manually:} Since we provide a Docker container with all dependencies pre-installed, the artifact relies on Docker and access to a web browser. Specific installation instructions can be found at \url{https://github.com/FPSG-UIUC/micro23-teaal-artifact/blob/main/README.md}.

\smallskip
\noindent \textbf{Via the MLCommons CM Interface:} The instructions for installation can be found at \url{https://github.com/ctuning/cm-reproduce-research-projects/blob/main/script/reproduce-ieee-acm-micro2023-paper-8/README.md}.

\subsection{Evaluation}

\textbf{Manually:} We provide two notebooks to guide you through the evaluation: \texttt{notebooks/figs9and10.ipynb} and \texttt{notebooks/fig11.ipynb}. Please launch the docker container, open the Jupyter Lab in a web browser, and navigate to this notebook.
Each cell either runs a simulation or displays a graph. The output of each display cell corresponds to a figure in the paper.

\smallskip
\noindent \textbf{Via the MLCommons CM Interface:} The instructions for evaluation can be found at \url{https://github.com/ctuning/cm-reproduce-research-projects/blob/main/script/reproduce-ieee-acm-micro2023-paper-8/README.md}

\subsection{Expected Results}

This artifact reproduces Figures~\ref{fig:eval:mem:extensor}-\ref{fig:eval:energy}. The easiest way to check validity is to visually compare the figures, but raw results will be written to \texttt{data/generated/} and can be compared with the expected results found in \texttt{data/pregenerated/}.
We note that certain experiments use randomly generated sparse tensors whose performance characteristics will exhibit some variety. Such datasets are noted in the notebook, and simulations can be rerun to obtain new seeds.\\

\subsection{Experiment Customization}

Input specifications in \texttt{yamls/teaal/} can be updated to work on new kernels, execute new mappings, represent tensors with new formats, and evaluate new architectures.\\


\subsection{Methodology}

Submission, reviewing and badging methodology:

\begin{itemize}
  \item \url{https://www.acm.org/publications/policies/artifact-review-and-badging-current}
  \item \url{http://cTuning.org/ae/submission-20201122.html}
  \item \url{http://cTuning.org/ae/reviewing-20201122.html}
\end{itemize}

\end{document}